\documentclass[notitlepage,aps,pra,superscriptaddress,showpacs,amsmath,amssymb,amsfonts,lengthcheck,twocolumn,longbibliography]{revtex4-1}

\usepackage{dsfont}
\usepackage{graphicx} 
\usepackage{graphics}

\usepackage{mathtools}
\usepackage{dcolumn}    
\usepackage{bm} 
\usepackage{graphicx}
\usepackage{amsmath}    
\usepackage{latexsym}
\usepackage{amsfonts}   
\usepackage{amssymb}
\usepackage{array}      
\usepackage{epsfig}
\usepackage{txfonts}
\usepackage{xcolor}
\usepackage[colorlinks=true,linkcolor=blue,urlcolor=blue,citecolor=blue,pdfusetitle]{hyperref}
\usepackage{hyperref}
\usepackage{ulem}

\newcommand{\ket}[1]{\left|#1\right\rangle}
\newcommand{\bra}[1]{\langle#1|}

\newcommand{\ketbra}[2]{|#1\rangle\langle#2|}

\newcommand{\avg}[1]{\langle#1\rangle}
\newcommand{\diag}{\text{diag}}

\newcommand{\Tr}{\mbox{Tr}}
\newcommand{\tr}{\mbox{Tr}}

\newcommand{\pket}[1]{\vert#1)}
\newcommand{\pbra}[1]{(#1\vert}

\newcommand{\pketbra}[2]{\vert #1 )(#2\vert}

\newcommand{\change}[1]{{\color{black}#1}}

\begin{document}
% \title{Extracting the detailed signal statistics of filtered continuously measured quantum processes}
\title{Extracting filtered signal statistics of continuously measured quantum systems}
% \title{Numerical Methods for Charge Resolved Master Equations}
%\title{Efficient simulation of weak delayed feedback control in quantum systems}
\author{Anthony Kiely}
\email{AKiely@ucc.ie}
\address{School of Physics, University College Dublin, Belfield Dublin 4, Ireland}
\affiliation{Centre for Quantum Engineering, Science, and Technology, University College Dublin, Ireland}
\affiliation{School of Physics, University College Cork, College Road, Cork, Ireland}
\author{Gabriel Landi}
\address{Department of Physics and Astronomy, University of Rochester, Rochester, New York 14627, USA}
\address{University of Rochester Center for Coherence and Quantum Science, Rochester, New York 14627, USA}

\begin{abstract}
% We show that the recently derived quantum Fokker-Planck master equation [Phys.
% Rev. Lett. 129, 050401 (2022)] for a continuously monitored and feedback controlled system, can be efficiently simulated. We present an efficient method for the steady state of the system and detector in the absence of feedback. 
The joint state of a continuously monitored quantum system and the classical filtered measurement record has recently been shown to be described by a quantum Fokker-Planck master equation [Phys. Rev. Lett. 129, 050401 (2022)]. We present a deterministic approach to compute the steady state of the system and detector. The method is shown to become particularly efficient in the absence of feedback, which we exploit to develop a perturbative approach valid for weak feedback. 
We show that through this method we can extract the full counting statistics of the signal,  the quantum-classical mutual information between system and signal, as well as the Fisher information of the signal, which can be used for sensing applications.
Our results are illustrated with both single-qubit models, as well as the spin chains governed by the one-dimensional transverse field Ising model or the Lipkin-Meshkov-Glick model.

% We use this to extract detailed statistics of the filtered signal, including all moments, the full probability distribution, the mutual information, correlation functions, as well as the Fisher information for parameter estimation. 
% % ADD A SENTENCE ABOUT EXTRACTING DETAILED STATISTICS: LIST WHAT WE CALCULATE: MOMENTS, FULL DISTRIBUTION, CORRELATION FUNCTIONS, ETC.
% % We use this to compute the Fisher information (a key quantity in quantum metrology) of the filtered measurement record. 
% Building on this, we demonstrate how perturbative corrections allow one to efficiently determine the steady state solution for weak feedback driving.
\end{abstract}

\maketitle

%\tableofcontents

\section{Introduction}

In contrast with textbook instantaneous projective measurements,  continuous measurements allow one to extract partial information about a quantum state without fully collapsing it~\cite{Landi2024}. 
In addition, continuous measurements allow for the implementation of feedback protocols, where the classical (noisy) data obtained from the measurements are fed back in real time to the system. 
Quantum continuous measurements have been experimentally realized across many quantum platforms, such as optical cavities~\cite{gleyzes2007}, quantum dots~\cite{gustavsson2006,bayer2025}, and superconducting circuits~\cite{vijay2011,murch2013,minev2019}.

The inclusion of feedback makes the dynamics non-linear and time non-local, dramatically complicating any theoretical description. Feedback schemes can be simulated numerically using stochastic trajectories. But deterministic descriptions remain extremely valuable. This is tantamount to the distinction between the Langevin equation and the Fokker-Planck equation in Brownian motion. The two descriptions are complementary. Both describe the same physical process, but from different perspectives.  
For quantum continuous measurements a similar distinction exists. 
In the absence of feedback, the analog of the Fokker-Planck equation (i.e., the deterministic evolution) is simply the quantum master equation. But in the presence of feedback this is no longer true. For this reason,
% From a theory perspective, these processes can be difficult to simulate, since they are described by stochastic master equations, which have to be sampled numerous times to extract the statistical behavior of the monitored system.
% This motivates the search for deterministic solution methods that can encompass the ensemble-averaged response of the system. 
traditionally, such methods have been largely restricted to what is known as current-based feedback, where at each step only the very last data point, called a current $z(t)$, is used to guide the feedback~\cite{wiseman2009}. 
In this paradigm the current $z(t)$ is a random variable and one could, for example, modify  the Hamiltonian based on the obtained value of $z$ at each step. 
More recently, this has been extended to charge-based feedback~\cite{Kewming2024}, where one uses instead the entire integrated current $N(t) = \int_0^t dt'~z(t')$. 

Both examples are actually particular cases of a low pass filter-based feedback, where one uses instead 
\begin{equation}\label{signal}
    D(t)=\int_{0}^t dt' \, \gamma e^{-\gamma(t-t')} z(t'),
\end{equation}
which recovers the current- and charge-resolved cases when $\gamma\to \infty$ and $\gamma\to 0$ respectively. 
Recently, a deterministic method for simulating feedback based on this kind of low pass filter was put forward in Ref.~\cite{annby-anderssonQuantumFokkerPlanckMaster2022}, for the case of the diffusive unraveling of the master equation.
In this formalism, in addition to the usual open system dynamics, an external agent is also continuously measuring a certain operator $A$ through a weak Gaussian measurement defined by Kraus operators
\begin{equation}
    K(z) = \left( \frac{2\lambda dt}{\pi}\right)^{1/4} e^{-\lambda dt(z-A)^2}, \label{eq:kraus}
\end{equation}
with strength $\lambda$. \change{In the limit of infinite measurement strength, this reduces to projective measurements of $A$.}
Ref.~\cite{annby-anderssonQuantumFokkerPlanckMaster2022} introduced a signal-resolved density matrix $\rho_t(D)$, defined such that $\Tr\big\{\rho_t(D)\big\} = P(D)$ equals the probability that the signal has a certain value $D$ at time $t$. \change{This mathematical object, $\rho_t(D)$, can be thought of as the joint (quantum-classical) state of the quantum system and the classical measurement apparatus.}
Their main result was to show that $\rho_t(D)$ satisfies a Fokker-Planck master equation of the form 
\begin{eqnarray}
\partial_t \rho_t(D) &=& \mathcal{L}(D)  \rho_t(D) + \lambda \mathcal{D}\left[A\right]  \rho_t(D)-\gamma \partial_D \mathcal{C}_{A-D} \rho_t(D)/2 \nonumber \\ &+&\frac{\gamma^2}{8 \lambda} \partial_D^2 \rho_t(D),\label{FP_eq}
\end{eqnarray}
where $\mathcal{L}(D)=\mathcal{L}_0 + \sum_p f_p (D) \mathcal{L}_p$ is the Liouvillian which can in principle have any dependence on the signal $D$.
Here $\mathcal{C}_X=\{X,\cdot\}$ is the anti-commutator with $X$ and $\mathcal{D}\left[A\right]=A \cdot A^\dagger - \frac{1}{2} \{A^\dagger A, \cdot\}$ is the Lindblad dissipator.
Feedback enters through the fact that the Liouvillian $\mathcal{L}(D)$ is allowed to depend on the signal $D$. And here we assume, without loss of generality, that this dependence enters through a set of functions $f_p(D)$ that couple to Liouvillian terms $\mathcal{L}_p$. This therefore allows for feedback that can be either coherent  (i.e., terms that enter into the system Hamiltonian) and incoherent (which enters into, e.g. jump operators). 
Eq.~\eqref{FP_eq} was derived assuming an evolution that combines the natural Liouvillian evolution, with $\mathcal{L}$, as well as the continuous application of Gaussian weak measurements defined by the Kraus operations in Eq. \eqref{eq:kraus}.
The new terms that appear on the right hand side of Eq.~\eqref{FP_eq} can be given a specific physical intuition:
% The first term describes the potentially feedback dependent Liouvillian. This term could also be time-dependent, although we do not present examples of this. 
the second term is the measurement back action, which leads to dephasing in the basis of the observable $A$ that one is measuring. 
And the third and fourth terms can be interpreted as analogous to the drift and diffusion terms in the standard Fokker-Planck equation.

%The measurement is performed  and the measurement outcomes experience a low pass frequency filter with a bandwidth $\gamma$.

% However, there has recently been an interest in using charge resolved master equations to avoid such difficulties . This involves considering different coarse graining of the measurement record.  One example of this was equations describing the first passage time statistics for the integrated measurement record . However similar master equations \cite{annby-anderssonQuantumFokkerPlanckMaster2022} were also derived earlier for low pass filtered measurement record, motivated by exploring feedback control for a delayed information signal.

This type of master equation has since been used to study the generation of steady-state entanglement from thermal resources using feedback \cite{Diotallevi_2024}, the cooling of a particle in a harmonic trap \cite{desousa2024} and the exploration of Maxwell demons in quantum and classical mechanics \cite{annby2024maxwell}.
Eq.~\eqref{FP_eq} has also recently been generalized to the case of multiple detectors \cite{desousa2024}.
However, actually solving Eq.~\eqref{FP_eq} is not an easy task. 
It has the structure of a Fokker-Planck equation (linear, second order parabolic differential equation). But the object itself is a density matrix. So this is akin to a system of coupled Fokker-Planck equations.
The starting point of this paper is a solution method briefly proposed in the Supplemental Material of Ref.~\cite{annby-anderssonQuantumFokkerPlanckMaster2022}.
Our first main result is the realization that, in the absence of feedback, this method can be modified to become extremely simple and efficient to use. 
It allows access to the full counting statistics of the signal $D(t)$ [Eq.~\eqref{signal}], including all statistical moments and the full probability density function $P_t(D)$. 
It also gives access to measures of correlation between system and signal; namely, the mutual information between $D(t)$  and the quantum state of the system, as well as any correlation function between $D(t)$ and a system observable. 
Finally, it gives access to the Fisher information contained in the signal, which is relevant when $D(t)$ is used for sensing applications. 
All of these results concern Eq.~\eqref{FP_eq} in the absence of feedback. 
We then move on to show that they can be extended to the case of weak feedback, using a perturbative approach which we develop. 
Taken together, our results therefore provide a comprehensive toolbox for efficiently extracting results from Eq.~\eqref{FP_eq}.
A brief comparison with stochastic simulations of quantum trajectories
is provided in Appendix~\ref{app:comparison_stochastic_trajs}.

\section{Formal framework}

% We are interested in finding the joint state of system and detector in the presence of feedback. In order to do this we will use three key steps. Firstly, we will write an explicit expression for the dynamical matrix which describes the FP equation in a convenient function basis. We then exploit the specific matrix structure to quickly determine the steady state under continuous monitoring without feedback. Finally, the state is reconstructed from these coefficient matrices using a reciprocal space to improve convergence.
% We can build on this numerical method for the steady state solution to determine the Fisher information of the filtered measurement current. We can also apply perturbative corrections to include the effects of weak feedback control.

% \subsection{Dynamical matrix}

Our starting point is a solution method of Eq.~\eqref{FP_eq} that was put forward in Ref.~\cite{annby-anderssonQuantumFokkerPlanckMaster2022}.
It consists in expanding the density operator in terms of a \change{complete basis of} generalized Hermite polynomials
\begin{eqnarray}
\rho_t(D) = \sum_{n=0}^{\infty} M_n \frac{\mathcal{H}_n(D)}{\sqrt{\sigma^n n!}} \frac{e^{-D^2/2\sigma}}{\sqrt{2 \pi \sigma}}, \label{ansatz}
\end{eqnarray}
with 
\begin{eqnarray}
\mathcal{H}_n(x) = \left(\frac{\sigma}{2}\right)^{n/2} H_n\left(\frac{x}{\sqrt{2 \sigma}}\right),
\end{eqnarray}
where $\mathcal{H}_n(x)$ are generalized Hermite polynomials,   $H_n(x)$ are standard physicist’s Hermite polynomials, and $M_n$ are Hermitian matrices of dimension $R$. 
Here we also defined the diffusion coefficient 
\begin{equation}
    \sigma = \frac{\gamma}{8\lambda}.
\end{equation}
% the diffusion constant $\sigma=\gamma/(8 \lambda)$ is the variance  
This choice of ansatz is motivated by the eigenfunctions of the drift and diffusion part of Eq. \eqref{FP_eq}. 
It means that we can fully describe the signal resolved state $\rho_t(D)$ in terms of the set of matrices $M_n$.
In principle, we need to consider all $M_n$ from $n=0$ to $\infty$. 
But, in practice, as shown below, the sum~\eqref{ansatz} can be truncated to a total of $N$ terms, where $N$ simply needs to be large enough to ensure convergence.
In the majority of the examples presented a value of $N\approx 100$ is found to be sufficient. For sufficiently large $N$, any state can be represented by this choice of ansatz.
The methods for actually extracting $M_n$ will be discussed starting in Sec.~\ref{sec:Mn}. In the remaining of this section, we assume that the $M_n$ are known, and show what kind of information can be extracted from them. We note, in particular, that for various quantities of interest only a handful of the $M_n$ matrices are actually needed.

% Before turning to how we can actually determine the matrices $M_n$, in the following subsections we show which quantitie}
% In the following parts, we will outline the different quantities that can be directly computed from these matrices, in some cases with only the first few. \change{We stress that several of these quantities such as the quantum-classical mutual information, and their connection to the matrices $M_n$ are novel.}

\subsection{Statistics of $D(t)$}

We start by defining the coefficients 
\begin{equation}\label{cm_def}
c_m=\Tr(M_m)= \int dD P(D)  \frac{\mathcal{H}_m(D)}{\sqrt{\sigma^m m!}}. 
\end{equation}
Note that $c_0=1$. 
The full distribution of the signal outcomes is obtained from Eq.~\eqref{ansatz} as 
\begin{equation}\label{PD_direct_reconstruction}
    P_t(D) = \Tr\big\{\rho_t(D)\big\} = \sum_{n=0}^\infty c_n \frac{\mathcal{H}_n(D)}{\sqrt{\sigma^n n!}} \frac{e^{-D^2/2\sigma}}{\sqrt{2 \pi \sigma}}.
\end{equation}
From this one can construct the statistical moments of the signal
\begin{equation}
    \avg{D^n}(t)=\int dD P_t(D) D^n.
\end{equation}
This is an ensemble expectation (i.e., taken over the ensemble of measurement outcomes). 
It can therefore be evaluated at any instant of time or in steady state.
Due to the orthogonality of the Hermite polynomials the average of $P_t(D)$ is 
\begin{equation}
\avg{D}= \sqrt{\sigma} c_1.
\end{equation}
Thus, if all one is interested is the average, it suffices to know $M_0$ and $M_1$ only.
Similarly, the second moment $\avg{D^2}= \sqrt{2} \sigma  c_2+\sigma$.  This gives the variance as 
\begin{equation}
    \rm{Var}(D)= \sigma (1+ \sqrt{2}  c_2  -  c_1^2).
\end{equation}
More generally, the $q^{\rm th}$ moment is
\begin{eqnarray}
    \avg{D^q} = \sum_n c_n J_n(q),
    \qquad 
    J_n(q)= \int dD D^q \frac{\mathcal{H}_n(D)}{\sqrt{\sigma^n n!}}  \frac{e^{-D^2/2\sigma}}{\sqrt{2 \pi \sigma}}. \nonumber \\
\end{eqnarray}
This integral can be solved recursively using
\begin{eqnarray}
    J_n(q) &=& \sqrt{\sigma (n+1)} J_{n+1}(q-1)+\sqrt{n \sigma} J_{n-1}(q-1), \\
    J_n(0) &=& \delta_{n,0}.
\end{eqnarray}
Since $J_n(q)$ is only nonzero for $n \leq q$, only the first $q+1$ coefficients are needed to compute $\avg{D^q}$. 
Thus, for computing $\avg{D^q}$ we only need to know $M_0,\ldots,M_q$.
% \change{These summary statistics provide a compact accessible description of the overall distribution of filtered measurement outcomes.}

% More generally, the $q^{\rm th}$ moment can be expressed as $\avg{D^q} = \sum_m a_m c_m$ where $a_m$ are given from the orthogonality relations as
% \begin{eqnarray}
%     a_n = \int dD D^q \frac{\mathcal{H}_n(D)}{\sqrt{\sigma^n n!}}  \frac{e^{-D^2/2\sigma}}{\sqrt{2 \pi \sigma}}.
% \end{eqnarray}
% \begin{eqnarray}
%     \avg{D^q} &=& \sum_{m=0} a_m c_m \\
%     &=&  \int dD P(D) \sum_{m=0}  a_m   \frac{\mathcal{H}_m(D)}{\sqrt{\sigma^m m!}},
% \end{eqnarray}

The characteristic function of $P_t(D)$ can also be constructed from the coefficients $c_n$ as
\begin{eqnarray}
    \avg{e^{i K D}}= e^{-K^2 \sigma/ 2} \sum_{n=0}^\infty c_n \frac{ (i K)^n \sigma^{n/2}}{\sqrt{n!}}.
\end{eqnarray}
From this we can then reconstruct $P_t(D)$ by taking the inverse Fourier transform. 
And, in practice, we have found this to be quite useful, since a direct reconstruction using Eq.~\eqref{PD_direct_reconstruction} leads to poor convergence due to fast oscillating terms. 
With the characteristic function $\avg{e^{i K D}}$, we can first  filter out very high frequencies, and then take the inverse Fourier transform.

\subsection{Quantum-classical correlations between system and detector}

The joint state of the system allows one to determine the correlations between the detector outcomes and the quantum system. First, one may define the quantum-classical mutual information between system and detector as,
\begin{eqnarray}
\mathcal{I}(t) = S \left[\int dD \rho_t(D) \right] + S \left[ \Tr\{\rho_t(D)\} \right]- S \left[ \rho_t(D) \right],
\end{eqnarray}
where $S \left[ \cdot \right]$ is either the von-Neumann or Shannon entropy depending on the input. This can be simplified (see Appendix \ref{app2}) as
\begin{eqnarray}\label{mutual_info_2}
    \mathcal{I}(t) =-\int dD \, S\left[ \tilde{\rho}_t(D) \right] P(D)+S[M_0],
\end{eqnarray}
where $\tilde{\rho}_t(D)=\rho_t(D)/P(D)$ is the conditional state of the system given an outcome $D$. \change{Eq.~\eqref{mutual_info_2} is a positive quantity, with a maximum value $\ln R$, where $R$ is the dimension of the Hilbert space of the system. Low values of $\mathcal{I}$ indicate that the system and detector are roughly statistically independent, while very high values correspond to a detector signal which is very informative about the system state.}

In general, computing $\mathcal{I}$ requires many matrices $M_n$ for convergence. A comparatively simpler quantity to calculate is the covariance between the signal and a particular system observable $B$, defined as
\begin{eqnarray}\label{covBD}
    \rm{cov}_t(B,D) &=& \int dD \, D \, \tr\big\{B \, \rho_t(D)\big\} - \avg{D} \avg{B} \\
    &=& \sqrt{\sigma} \left[ \tr(M_1 B) -  \tr(M_1) \tr (M_0 B) \right],
\end{eqnarray}
where in the second line (derived in Appendix \ref{app2}), we have shown how this can be computed using only $M_0$ and $M_1$. \change{Intuitively, $\rm{cov}(B,D)$ provides a measure of the correlation between the quantum state in the basis of the observable $B$ and the continuous filtered measurement outcome D. Note that the usual quantum mutual information is lower bounded by the corresponding covariance \cite{Wolf2008}, motivating us to consider the analogous pair of objects in the context of continuous monitoring.}
% This covariance lower bounds the mutual information by a quantum-classical Pinsker's inequality
% \begin{eqnarray}
%     \frac{\left|\rm{cov}(B)\right|}{||B||} \leq \sqrt{2 I_c}\leq \sqrt{2 I_q}.
% \end{eqnarray}

\subsection{Two-time correlation and Fisher information of steady state filtered output current}

The quantities described above can be determined solely from knowledge of the matrices $M_n$, in cases with or without feedback.
In this section we describe two other quantities of interest, whose calculation is somewhat more complicated. In particular, we have only been able to find explicit formulas for them in the case of no feedback. Their actual computation is done in Sec.~\ref{ssec:no_feedback}. 
Eq.~\eqref{covBD} is an equal time correlation function. One can also be interested in two-time correlation functions. 
The two-time function of the signal is defined as 
\begin{equation}
     C(t,\tau) = \avg{D(t+\tau)D(t)}-\avg{D(t)}^2.
\end{equation}
In Eq.~\eqref{two_time_corr} below and Appendix~\ref{app3}, we provide an explicit formula for this quantity in the case of no feedback, and in the steady-state (for which $C(t,\tau)$ is a function only of $\tau$).

Next, we turn to the Fisher information. Let $\mu$ denote any parameter that appears in Eq.~\eqref{FP_eq}. This could be some term in the Hamiltonian or some jump operator. 
The probability distribution $P_t(D)$ contains information about $\mu$. And therefore measurements of $D$ can be used to estimate $\mu$. 
The Fisher information represents the total information about $\mu$ that is contained in $P_t(D)$, and is defined as 
\begin{eqnarray}\label{fisher_info}
    F_\mu &=& \int dD \frac{\big[\partial_\mu P_t(D)\big]^2}{P_t(D)}.
\end{eqnarray}
The Fisher information lower bounds the variance of the estimated parameter $\mu$ via the Cram\'er-Rao bound \cite{Cramer1946,Rao1945}.
We note that $F_\mu$ refers to the classical Fisher information of the empirical distribution \cite{smiga2023}, where the temporal correlations are discounted, i.e. readings are collated to build a histogram for $P(D)$ in steady state. In practice, we must compute the Fisher information at some fixed value of the parameter. The numerical computation of $\partial_\mu P_\mu (D)$  will be addressed in the next section, for the setting of the no feedback steady-state.

\section{Evolution equation for $M_n$}\label{sec:Mn}

The results in the previous section are all based on the decomposition~\eqref{ansatz} of the signal-resolved density matrix, which is defined by the set of matrices $M_n$. 
We now turn to the  question of how to actually compute these $M_n$, which is based on the method proposed in~\cite{annby-anderssonQuantumFokkerPlanckMaster2022}.
We start by using the decomposition $\mathcal{L}(D)=\mathcal{L}_0 + \sum_p f_p (D) \mathcal{L}_p$ of the Liouvillian, into the feedback independent part $\mathcal{L}_0$ and the feedback terms $\mathcal{L}_p$.
Inserting Eq.~\eqref{ansatz} in Eq.~\eqref{FP_eq} yields a system of coupled equations for the operators $M_n$ (see Appendix \ref{app} for more details): 
\begin{equation}\label{FPbasis}
\begin{aligned}
    \dot{M}_m =& \mathcal{L}_0 M_m + \sum_{n=0}^{N-1}  \sum_p \alpha_{n,m,p} \mathcal{L}_p M_n -\frac{\lambda}{2}\left[A,[A,M_m]\right] 
    \\[0.2cm]
    &-\frac{\gamma}{2} \left[ 2 m M_m-\sqrt{\frac{m}{\sigma}}   \{A,M_{m-1}\}  \right], 
\end{aligned}    
\end{equation}
where
\begin{eqnarray}
    \alpha_{n,m,p} = \int_{-\infty}^\infty dD \frac{\mathcal{H}_m(D)}{\sqrt{\sigma^m m!}}  \frac{\mathcal{H}_n(D)}{\sqrt{\sigma^n n!}} \frac{e^{-D^2/2\sigma}}{\sqrt{2 \pi \sigma}}  f_p(D),
\end{eqnarray}
are coefficients determined by the feedback functions $f_p(D)$.
For example, in the case of a Heaviside step function $f_p(D)=\theta(D)$, these coefficients become~\cite{annby-anderssonQuantumFokkerPlanckMaster2022}
\begin{eqnarray}
\alpha_{n,m,p}
%&=& \int_0^\infty \frac{\mathcal{H}_m(x)}{\sqrt{\sigma^m m!}} \frac{\mathcal{H}_n(x)}{\sqrt{\sigma^n n!}} \frac{e^{-x^2/2\sigma}}{\sqrt{2 \pi \sigma}} dx \nonumber \\
&=& \begin{cases} 
      1/2 & n=m \\
      0 & n+m \, \, \rm{even} \\
      \frac{(-1)^{(n+m-1)/2} m!!(n-1)!!}{\sqrt{2 \pi n! m!} (m-n)}
      % C_{n,m} 
      & n+m  \,\, \rm{odd}
   \end{cases}.
\end{eqnarray}
% where
% \begin{eqnarray}
% C_{n,m}=\frac{(-1)^{(n+m-1)/2} m!!(n-1)!!}{\sqrt{2 \pi n! m!} (m-n)}.
% \end{eqnarray}
Similarly in the case of linear feedback $f_p(D)=D$, we get 
\begin{eqnarray}
\alpha_{n,m,p} &=&    \sqrt{\sigma } \left(\sqrt{n+1} \delta
   _{m-1,n}+\sqrt{n} \delta _{m+1,n}\right).
\end{eqnarray}
The matrix elements can be computed analytically for any case where $f_p(D)$ is polynomial in $D$ using recursive relations, as summarized in Appendix \ref{app0}. 
%If $f_p(D)$ has no $D$ dependence, then $\alpha_p = 1_N$.

Eq.~\eqref{FPbasis} can be written more compactly as a single vector equation. We truncate the infinite sum in Eq.~\eqref{ansatz} to a sufficiently large $N$ and define a vector of matrices  ${\vec M} = \left( M_0 \, M_1 \, M_2 \ldots M_{N-1} \right)^T$. Our dynamical equation can now be succinctly expressed as
\begin{equation}\label{Mvec_eq}
 \partial_t \vec{M} = \hat{Q} \vec{M},  
\end{equation}
where $\hat{Q}=\hat{Q}_0 + \hat{Q}_{\rm fb}$ with
\begin{eqnarray}
    \hat{Q}_0 &=& 1_N \otimes \mathcal{L}_0 + \lambda 1_N \otimes \mathcal{D}[A] -\gamma F \otimes 1_d + \frac{\gamma}{2 \sqrt{\sigma}} G \otimes \mathcal{C}_A, \nonumber \\
    \hat{Q}_{\rm fb} &=&  \sum_p \alpha_p \otimes \mathcal{L}_p.
\end{eqnarray}
% \begin{eqnarray}
%     \hat{Q} &=& 1_N \otimes \mathcal{L}_0 + \sum_p \alpha_p \otimes \mathcal{L}_p + \lambda 1_N \otimes \mathcal{D}[A] \nonumber \\
%     &-& \gamma F \otimes 1_d + \frac{\gamma}{2 \sqrt{\sigma}} G \otimes \mathcal{C}_A, \label{qmatrix}
% \end{eqnarray}
Here $1_x$ is the identity operator of dimension $x$, whereas $F$, $G$ and $\alpha_p$ are $N\times N$ matrices with elements 
\begin{align}
    F&=\diag(0,1,2,\ldots, N-1),
    \\[0.2cm]
    \left[G\right]_{n,m}&=\sqrt{j} \,\delta_{n,m+1},
    \\[0.2cm]
    \left[\alpha_p\right]_{n,m}&=\alpha_{n,m,p}.   
\end{align}
The system Hilbert space has dimension $R$, so the Liouvillian has dimension $d=R^2$.
The total dimension of the matrix $\hat{Q}$ is then $R^2 N$.

\subsection{The case of no feedback}\label{ssec:no_feedback}

The challenging aspect of Eq.~\eqref{FPbasis}, or its more compact version~\eqref{Mvec_eq}, is that they involve a set of coupled equations for all the matrices $M_m$ which must therefore be solved simultaneously.
An interesting realization, however, is that the entire result simplifies dramatically if there is no feedback.
That is, if all functions $f_p(D) \equiv 0$. 
This implies the matrices $\alpha_p=0$ vanish, and so only $\hat{Q}_0$ contributes to Eq.~\eqref{Mvec_eq}. 
This operator, however, has the very special property of being block lower triangular, which means that each $M_m$ depends only on the previous one. 
% \anto{For the case of no feedback control, the matrix  has a specific lower triangular form. Firstly, this implies that the characteristic polynomial is a product of the characteristic polynomial for each block on the main diagonal. The eigenvalues are then a union of the eigenvalues of each of these blocks. Each block has eigenvalues with negative real part and come in complex conjugate pairs. The spectrum of $\hat{Q}_0$ then inherits these properties.}
For the steady-state solution ($\dot{M}_m=0$), all of the matrices can therefore be solved using forward substitution. 
Namely, we first solve for $M_0$. Then we insert this into the equation for $m=1$ and solve for $M_1$. This, in turn, is inserted into the equation for $m=2$ which yields $M_2$. This can then be continued in an iterative fashion. 
This matches well with the fact, discussed in the previous section, that various quantities of interest only require a handful of $M_m$.
% The steady state (i.e. the eigenvector associated with the zero eigenvalue) can be solve using forward substitution as follows.  
The equation for $M_0$ becomes 
\begin{equation}\label{M0_eq}
    \Lambda M_0=0,
    \qquad 
    \Lambda := \mathcal{L}_0+ \lambda \mathcal{D}[A],
\end{equation}
% First one solves $\Lambda M_0=0$ for $M_0$ where $\Lambda=\mathcal{L}_0+ \lambda \mathcal{D}[A]$.  Note that this 
which is simply the usual steady state solution for the unconditional dynamics, with Liouvillian $\Lambda$. From $M_0$ one can then recursively solve for the other terms using 
\begin{eqnarray}
(\Lambda-\gamma n 1_d) M_n =  -\frac{\gamma}{2} \sqrt{\frac{n}{\sigma}} \mathcal{C}_A \left( M_{n-1} \right), \label{update}
\end{eqnarray}
for $n=1,2,\ldots$. Note that provided there is a unique solution for $M_0$, this solution is uniquely determined. 
This iterative process can also be made faster by first finding the spectrum of the unconditional Liouvillian
\begin{equation}
\Lambda= \sum_{j \neq 0} \eta_j \pketbra{x_j}{y_j} ,    
\end{equation}
where $\pket{x_j}$ and $\pbra{y_j}$ are the right and left eigenvectors associated with eigenvalue $\eta_j$. Specifically $\pket{x_0}=\pket{M_0}$, $\pbra{y_0}=\pbra{1_d}$ and $\eta_0=0$. 
Even though a Liouvillian itself is not invertible, the quantity $\Lambda-\gamma n 1_d$ will be, for any $n \neq 0$. In fact, from the eigenspectrum on readily finds
\begin{eqnarray}
    (\Lambda-\gamma n 1_d)^{-1} = \sum_{j \neq 0} (\eta_j- \gamma n)^{-1} \pketbra{x_j}{y_j} - \frac{1}{\gamma n} \pketbra{M_0}{1_d}.
\end{eqnarray}
The recursion relation can then be expressed as
\begin{eqnarray}
    \pket{M_n} &=& - \frac{\gamma}{2} \sqrt{\frac{n}{\sigma}}  \sum_{j = 0} \frac{\pbra{y_j} \mathcal{C}_A \pket{M_{n-1}}}{\eta_j- \gamma n}  \pket{x_j}.
\end{eqnarray}
% assuming $\eta_j-\gamma n \neq 0$ for all $j$ and $n$.
% This provides a simple analytical expression for the matrices as
% \begin{eqnarray}
%     M_n=- \frac{\gamma}{2} \sqrt{\frac{n}{\sigma}} \sum_{j=0} \frac{\Tr\left[y_j^\dagger \{A, M_{n-1}\}\right]}{\eta_j- \gamma n}  x_j,
% \end{eqnarray}
% where we have assumed $\eta_j-\gamma n \neq 0$ for all $j$ and $n$.

This same decomposition also allows one to compute the correlations in the measurement current (see Appendix \ref{app3}),
\begin{eqnarray}\label{two_time_corr}
    C(\tau) &=& \avg{D(t+\tau)D(t)}-\avg{D(t)}^2\\
     &=&\sigma e^{-\gamma \tau} + \frac{1}{2} \sum_{j \neq 0}  \frac{\gamma  \left(\gamma  e^{\tau  \eta
   _j}+e^{-\gamma  \tau } \eta _j\right)}{\gamma
   ^2-\eta _j^2} \Tr(A x_j) \Tr(y_j^\dagger \{A,M_0\}). \nonumber \\
\end{eqnarray}
\change{Since the measurement current is typically highly correlated in time, this measure of the fluctuations provides information not accessible from the average, see for example \cite{Landi2024}. This is also a highly accessible quantity experimentally.}

% \subsection{Parameter derivatives}\label{ssec:parameter_derivatives}

Finally, we can also exploit the recursive nature of Eq.~\eqref{update} to obtain an expression for evaluating the derivative $\partial_\mu P_t(D)$ of the signal probability distribution with respect to an arbitrary parameter $\mu$ appearing in Eq.~\eqref{FP_eq}. This quantity is necessary to compute the Fisher information defined in Eq.~\eqref{fisher_info}.
A straightforward method for computing $\partial_\mu P_t(D)$ is to solve for $P_t(D)$ for a certain value $\mu$ and a nearby value $\mu+\delta\mu$, from which $\partial_\mu P_t(D)$ can then be computed numerically using finite differences. 
However, $\delta \mu$ must be sufficiently small for the derivative to be well approximated numerically, but not so small as to become comparable with the other numerical errors in the procedure. Due to this non-trivial interplay, we have actually found this method to be numerically quite inaccurate.
Instead, we now provide another method, which we have found to be quite efficient, since it completely avoids finite differences. 
It only works in the absence of feedback and in the steady-state. And it assumes that $\mu$ is neither $\lambda$ nor $\gamma$.
From Eqs.~\eqref{cm_def} and~\eqref{PD_direct_reconstruction} it follows that 
\begin{equation}
    \partial_\mu P_t(D) = \sum_{n=0}^\infty \Tr\big\{ \partial_\mu M_m\big\} \frac{\mathcal{H}_n(D)}{\sqrt{\sigma^n n!}} \frac{e^{-D^2/2\sigma}}{\sqrt{2 \pi \sigma}}.
\end{equation}
Hence, what we ultimately need is $\partial_\mu M_m$.
% When the state $M_n^\mu$ depends on a system parameter (described by $\mathcal{L}_0^\mu$), one can continue to use forward substitution (in the case of no feedback) as before to efficiently compute $P_\mu(D)$.
% However, to calculate the Fisher information of the measurement signal, we must extend this to also compute $\partial_\mu M_n^\mu$ for the derivative term $\partial_\mu P_\mu(D)$. 
We start with $M_0$. 
Differentiating Eq.~\eqref{M0_eq} with respect to $\mu$ yields
% Taking the derivative of $\Lambda^\mu M_0^\mu =0$ (where $\Lambda^\mu=\mathcal{L}_0^\mu + \lambda \mathcal{D}[A]$), we can rewrite this as 
\begin{equation}
    \Lambda (\partial_\mu M_0) = - (\partial_\mu \mathcal{L}_0) M_0,
\end{equation}
which can be solved for $\partial_\mu M_0$. 
To get the next terms, we can use the iterative update rule in Eq. \eqref{update}. Differentiating both sides with respect to $\mu$, provides the modified update rule
\begin{eqnarray}
\left(\Lambda-\gamma n 1_d \right) \left( \partial_\mu M_n \right)=-\left(\partial_\mu \mathcal{L}_0\right) M_n  -\frac{\gamma}{2} \sqrt{\frac{n}{\sigma}} \mathcal{C}_A \left( \partial_\mu M_{n-1} \right), \nonumber \\
\end{eqnarray}
so given $\mathcal{L}_0$, $\partial_\mu M_{n-1}$ and $M_n$ we can once again iteratively solve for $\partial_\mu M_n$.
This method is simple to implement because the derivative $\partial_\mu \mathcal{L}_0$ is simple to calculate analytically.

All of the results in this subsection assume no feedback.
At first this might seem to defeat the entire point, since Eq.~\eqref{FP_eq} was actually derived precisely to describe feedback. 
However, we note two important points. 
First,  as we show next, the results in this section can actually be used as the starting point for a perturbative solution in the case when the feedback is weak. This is a strong motivation in itself.
Second, the case of no feedback is actually still of interest, since this is precisely what is studied in the field of Full Counting Statistics~\cite{Landi2024}. 
The only new ingredient is the presence of a low-pass filter, which is reasonable from an experimental point of view. 
This can also be removed, i.e., in the limit $\gamma \to 0$ the expressions presented in this section recover the standard results of full counting statistics~\cite{Landi2024}.
Quantities such as the signal distribution $P_t(D)$ are therefore still valuable and interesting, even in the absence of feedback.

\subsection{Perturbative solution for weak feedback}

\change{In the case of feedback, one must solve the full equation~\eqref{Mvec_eq} for the vector of matrices ${\vec M}$. But because the dynamical matrix $\hat{Q}$ does not have the block lower triangular structure of $\hat{Q}_0$, the previous approach based on forward substitution cannot be applied. Instead, one must solve for the entire ${\vec M}$ at once, which is computationally much more costly.}
The situation simplifies, however, if we consider a weak feedback of the form $\hat{Q}=\hat{Q}_0 + \epsilon \hat{Q}_{\rm fb}$, with $\epsilon \ll 1$.
We focus here on the steady-state of Eq.~\eqref{Mvec_eq}, which is a solution of 
\begin{equation}
    \hat{Q} {\vec M} = 0.
\end{equation}
Series expanding the solution as $\vec{M} \approx \sum_j \epsilon^{j} \vec{M}^{(j)}$ yields another recursive set of relations 
\begin{eqnarray}
  \hat{Q}_0 \vec{M}^{(0)} &=& 0,  
    \\[0.2cm]
  \hat{Q}_0 \vec{M}^{(j+1)} &=&  - \hat{Q}_{\rm fb} \vec{M}^{(j)},
\end{eqnarray}
for $j = 0,1,2,\ldots$ . More explicitly, what we find is the following generalization of Eqs.~\eqref{M0_eq} and~\eqref{update}:
\begin{align}
    \Lambda M_0^{(j+1)} &= -\big[ \hat{Q}_{\rm fb} \vec{M}^{(j)}\big]_0,
    \\[0.2cm]
    (\Lambda-\gamma n 1_d)M_n^{(j+1)}&= -\frac{\gamma}{2} \sqrt{\frac{n}{\sigma}} \mathcal{C}_A \left( M_{n-1}^{(j+1)} \right) - \big[ \hat{Q}_{\rm fb} \vec{M}^{(j)}\big]_n,
\end{align}
with $n = 1,2,\ldots$ and $j = 0,1,2,\ldots$. 
We first solve these for $j=0$ and all $n$ up until the truncation $N$. Then we plug these results and solve for all matrices with $j=1$, and so on, and so forth.
Because $\hat{Q}_0$ is lower triangular, solving this system at each iteration is roughly as costly as the case with no feedback.

\section{Numerical examples}

\subsection{Driven qubit}

\begin{figure}[t]
\centering
\includegraphics[width=\linewidth]{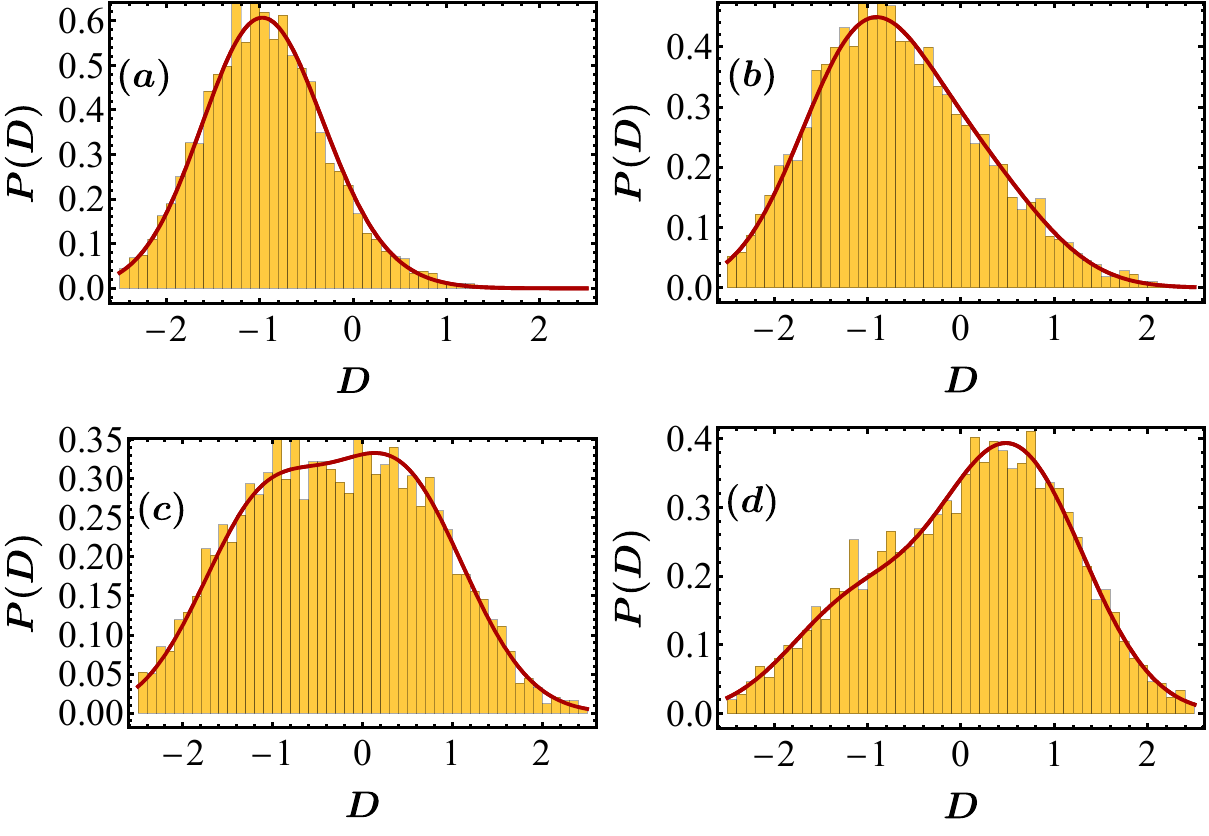}
\caption{Measurement outcomes $P(D)$ solving charge resolved master equation (solid red line) and collating 5000 stochastic trajectories (orange histograms) for $\Omega=1$, $\lambda=0.5$, $\gamma=2$ and $A=\sigma_z$ at different times (a) $t=\pi/8$ (b) $t=\pi/4$ (c) $t=3 \pi/8$ and (d) $t=\pi/2$.} 
\label{fig:compare}
\end{figure}

We illustrate the basic ideas with the simplest Hamiltonian $H=\Omega \sigma_x$ with a measurement $A=\sigma_z$. In Fig. \ref{fig:compare}, we can see a comparison between the charge resolved approach in red and the statistical collating of trajectories in the orange histograms. We see good agreement between the two at different times during the Rabi oscillations. \change{A more detailed comparison with the method of stochastic trajectories can be found in Appendix \ref{app:comparison_stochastic_trajs}.}

\begin{figure}[t]
\centering
\includegraphics[width=0.49\linewidth]{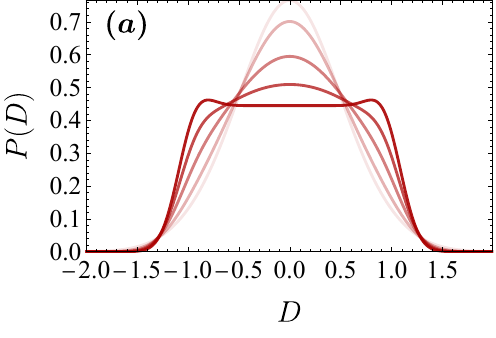}
\includegraphics[width=0.49\linewidth]{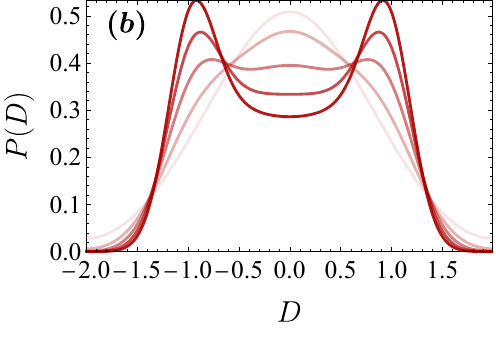}
\caption{Steady state distribution of detector outcomes $P(D)$ for increasing values of measurement strength $\lambda=\{0.5,1,1.5,2,2.5\}$ for darker shaded lines with measurement bandwidth (a) $\gamma=0.5$ and (b) $\gamma=1$.}
\label{fig:QubitSS}
\end{figure}

Our approach allows us to analytically calculate the summary statistics of the steady state probability distribution $P(D)$. In particular, for this example we get that $\avg{D}=0$ and $\rm{Var}(D)=\frac{\gamma  (\gamma +2 \lambda )}{\gamma ^2+2 \gamma  \lambda +4 \Omega^2}+\frac{\gamma }{8 \lambda}$. The steady state distribution is shown in Fig. \ref{fig:QubitSS}. Note that for increasing measurement strength the outcomes become more localized around $\pm 1$, the eigenvalues of $A$. The value of $\gamma$ is associated with the broadening of these peaks.

% We consider a system with a Hamiltonian $H_\mu=(\Omega+ \mu ) \sigma_x$, which enters as $\mathcal{L}^\mu_0=-i [H_\mu,\cdot]$. We perform a measurement $A=\sigma_z$ to estimate the correction to a known reference Rabi frequency $\Omega$. For this particular case, we can analytically calculate the summary statistics of the steady state probability distribution $P_\mu(D)$ which encodes information about the unknown parameter $\mu$. In particular we have that $\avg{D}=0$ and $\rm{Var}(D)=\frac{\gamma  (\gamma +2 \lambda )}{\gamma ^2+2 \gamma  \lambda +4 (\mu +\Omega )^2}+\frac{\gamma }{8 \lambda }$.

%%%%%%%%%%%%%%%%%%%%%%%%%%
%%%%%%%%%%%%%%%%%%%%%%%%%%

%%%%%%%%%%%%%%%%%%%%%%%%%%
%%%%%%%%%%%%%%%%%%%%%%%%%%

\begin{figure}[t]
\centering
\includegraphics[width=0.49\linewidth]{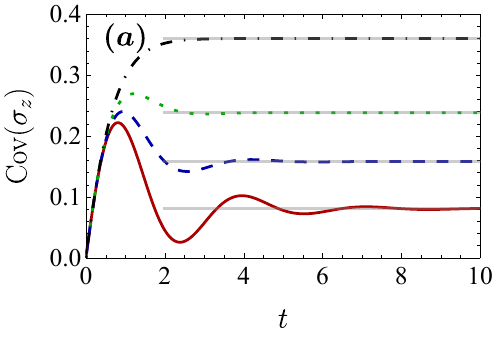}
\includegraphics[width=0.49\linewidth]{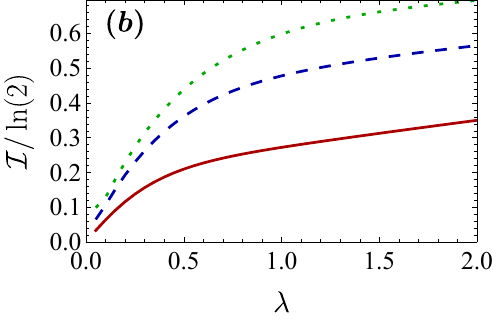} \\
\includegraphics[width=0.49\linewidth]{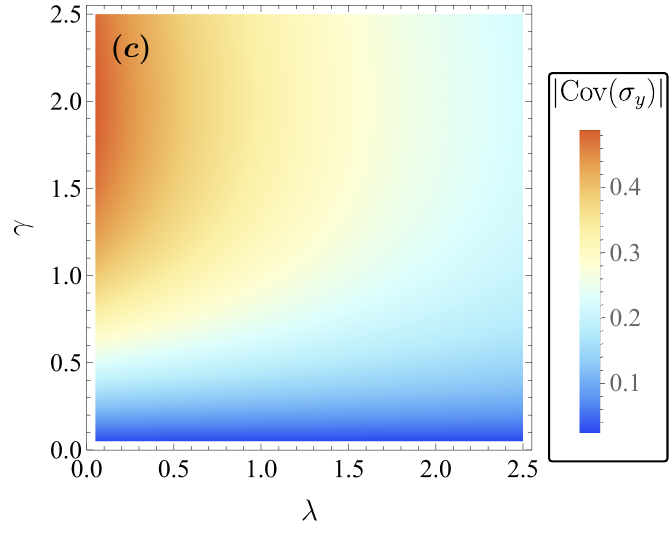}
\includegraphics[width=0.49\linewidth]{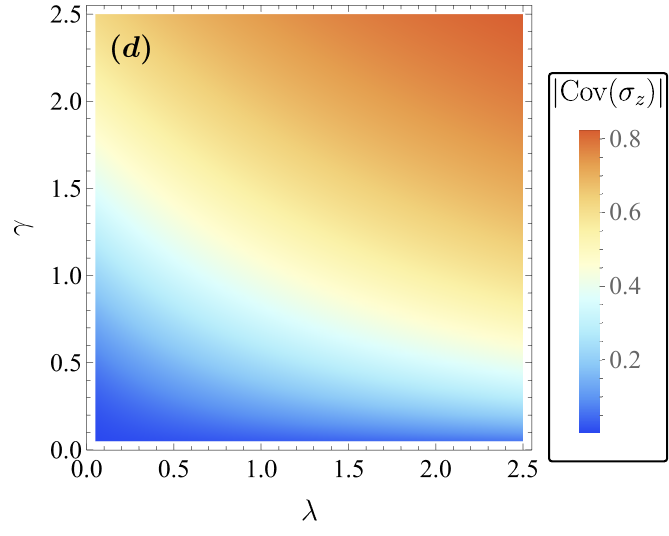}
\caption{ System detector correlations for the driven qubit. (a) Evolution of $\rm{Cov}(\sigma_z)$ with time for $\lambda=\{0.1,0.5,1.0,2.0\}$ (red solid, blue dashed, green dotted and black dot dashed lines). Steady state values are shown by faded gray lines. (b) Steady state mutual information $\mathcal{I}$ against measurement strength $\lambda$ for different bandwidths $\gamma=\{0.5,1,1.5\}$(red solid, blue dashed and green dotted lines). (c) Steady state covariance for $\sigma_y$ and (d) $\sigma_z$. Other parameter values: $\Omega=1$, $\gamma=0.5$. 
} 
\label{fig:DynamcicsZ}
\end{figure}

The dynamical accumulation of correlations between the system and the detector can be seen in Fig. \ref{fig:DynamcicsZ}(a). The covariance initially grows with time, eventually matching the steady state value (faded gray lines). The steady state covariance is also clearly larger for increasing values of measurement strength, which can also be seen in the analytical result,
\begin{eqnarray}
  \rm{Cov}(\sigma_z)  =\frac{\gamma  (\gamma +2 \lambda )}{\gamma ^2+2 \gamma  \lambda +4 \Omega ^2}.
\end{eqnarray}

The steady state mutual information is shown in Fig. \ref{fig:DynamcicsZ}(b), which again increases with increasing measurement strength $\lambda$ and bandwidth $\gamma$.
The steady state covariance for $\sigma_y$ and $\sigma_z$ is shown in Fig. \ref{fig:DynamcicsZ}(c,d) as a function of measurement strength $\lambda$ and bandwidth $\gamma$. We exclude $\sigma_x$ since $\rm{Cov}(\sigma_x)=0$. Since the measured observable is $A=\sigma_z$, there is a much higher correlation in the measurement signal and $\sigma_z$ than $\sigma_y$. $|{\rm Cov}(\sigma_z)|$ also increases with both $\lambda$ and $\gamma$.

%%%%%%%%%%%%%%%%%%%%%%%%%%
%%%%%%%%%%%%%%%%%%%%%%%%%%

% %%%%%%%%%%%%%%%%%%%%%%%%%%
% \begin{figure}[t]
% \centering
% \includegraphics[width=0.49\linewidth]{DynamcicsZ.pdf}
% \includegraphics[width=0.49\linewidth]{MI_bound.pdf}
% \caption{(a) Evolution of $\rm{Cov}(\sigma_z)$ with time for $\lambda=0.1,0.5,1.0,2.0$ (red solid, blue dashed, green dotted and black dot dashed lines). Steady state values are shown by faded gray lines. (b) Steady state mutual information $\mathcal{I}$ (solid red line) against measurement strength with lower bounds from the covariances of $\sigma_x$ (blue dashed line), $\sigma_y$(green dotted line) and $\sigma_z$ (black dot dashed line). Other parameter values: $\Omega=1$, $\gamma=0.5$. 
% } 
% \label{fig:DynamcicsZ}
% \end{figure}
% %%%%%%%%%%%%%%%%%%%%%%%%%%

% %%%%%%%%%%%%%%%%%%%%%%%%%%
% \begin{figure}[t]
% \centering
% \includegraphics[width=0.49\linewidth]{qubitdensityY.pdf}
% \includegraphics[width=0.49\linewidth]{qubitdensityZ.pdf}
% \caption{Steady state covariance for the driven qubit with $\Omega=1$.} 
% \label{fig:SScovariance}
% \end{figure}
% %%%%%%%%%%%%%%%%%%%%%%%%%%

%%%%%%%%%%%%%%%%%%%%%%%%%%
% \begin{figure}[t]
% \centering
% \includegraphics[width=0.8\linewidth]{MI_bound.pdf}
% \caption{Steady state mutual information $\mathcal{I}$ against measurement strength with lower bounds from the covariances of $\sigma_x$ (blue), $\sigma_y$(green) and $\sigma_z$ (black). $\Omega=1$ and $\gamma=0.5$} 
% \label{fig:MI_bound}
% \end{figure}
%%%%%%%%%%%%%%%%%%%%%%%%%%

\begin{figure}[t]
\centering
\includegraphics[width=0.49\linewidth]{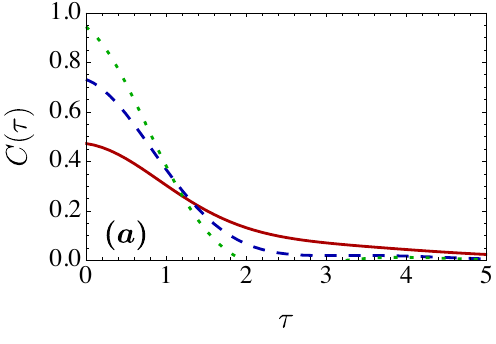}
\includegraphics[width=0.49\linewidth]{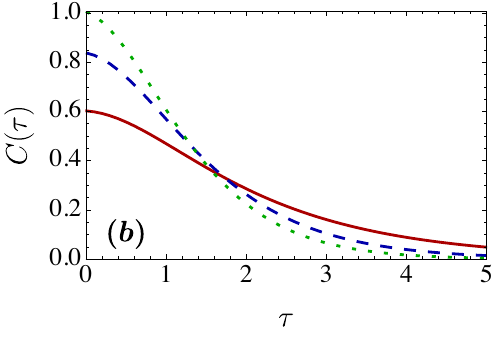}
\caption{Two time correlation function for the measurement current $C(\tau)$ with parameters 
$\Omega=1$ and  $\gamma=\{0.6,1,1.4\}$(solid red, dashed blue line and dotted green line) for measurement strengths (a) $\lambda=1$ and (b) $\lambda=1.9$.} 
\label{fig:CurrentCorrelationQubit}
\end{figure}

The two-time measurement current correlation can also be easily computed in the absence of feedback. For this example, we can see this in Fig. \ref{fig:CurrentCorrelationQubit}. For increasing values of $\gamma$, the correlations decay in time more dramatically.
%%%%%%%%%%%%%%%%%%%%%%%%%%
%%%%%%%%%%%%%%%%%%%%%%%%%%

%%%%%%%%%%%%%%%%%%%%%%%%%%
\begin{figure}[t]
\centering
\includegraphics[width=\linewidth]{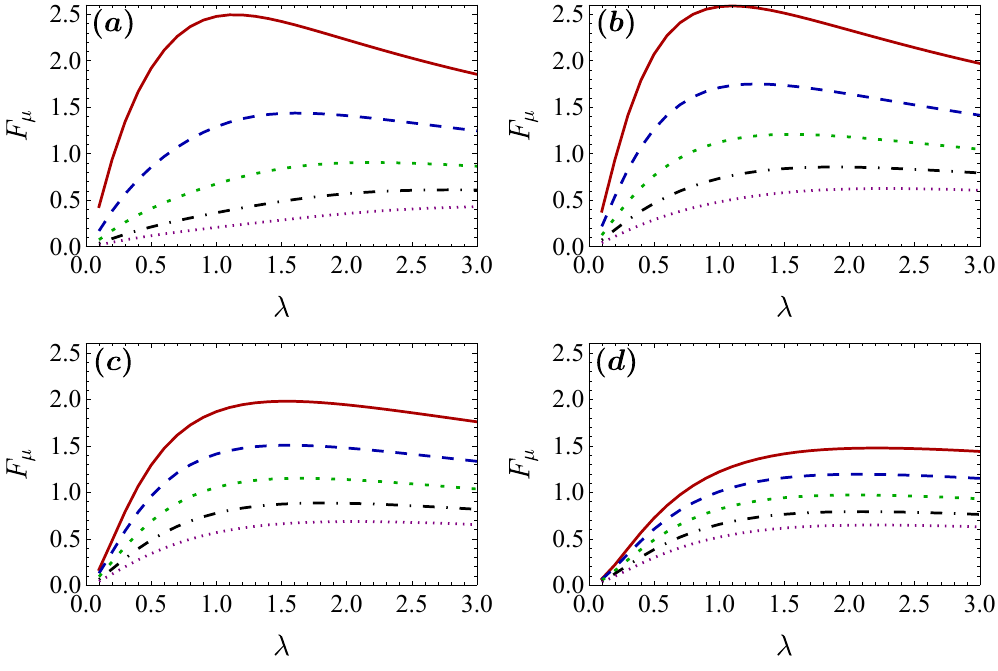}
\caption{Fisher information $F_\mu$ of the steady state distribution $P_\mu(D)$ against measurement strength $\lambda$ for different bandwidths $\gamma=\{0.8,1,1.2,1.4,1.6\}$ (solid red, dashed blue, dotted green, dashed black and short dashed purple lines). Each panel shows results for different values of the reference Rabi frequency (a) $\Omega=0.2$, (b) $\Omega=0.4$, (c) $\Omega=0.6$, and (d) $\Omega=0.8$.} 
\label{fig:qubitFI}
\end{figure}
%%%%%%%%%%%%%%%%%%%%%%%%%%

One could also estimate a correction to the known reference Rabi frequency as $H_\mu=(\Omega+ \mu ) \sigma_x$, which enters as $\mathcal{L}^\mu_0=-i [H_\mu,\cdot]$. In Fig. \ref{fig:qubitFI} we show the Fisher information for a range of parameters. First of all it is clear that for no measurement ($\lambda=0$) one gets no information $F_\mu=0$. The Fisher information also vanishes in the limit of very large measurement strength due to the quantum Zeno effect \cite{Facchi2008} where the dynamics is frozen. Another point worth noting is that the Fisher information decreases with increasing bandwidth $\gamma$. This is because a large $\gamma$ corresponds to the $D(t)$ being the instantaneous measurement outcome, while small $\gamma$ approaches the case where $D(t)$ is the full integrated current.

\subsection{Multi-qubit models}
In order to showcase how our method can tackle systems with multiple qubits, we will consider both the Ising and Lipkin-Meskov-Glick (LMG) models.

The Ising Hamiltonian is 
\begin{eqnarray}
    H=J \sum_{j=1}^{L-1} \sigma_{j+1}^z \sigma_j^z + h \sum_{i=1}^L \sigma_i^x,
\end{eqnarray}
where $L$ is the number of qubits. We will consider a measurement of the total magnetization $A=\sum_{i=1}^L \sigma_i^z$.

\begin{figure}[t]
\centering
\includegraphics[width=0.49\linewidth]{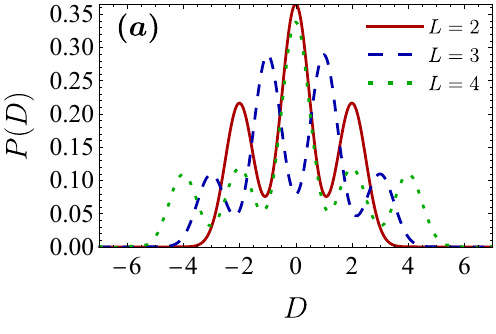}
\includegraphics[width=0.49\linewidth]{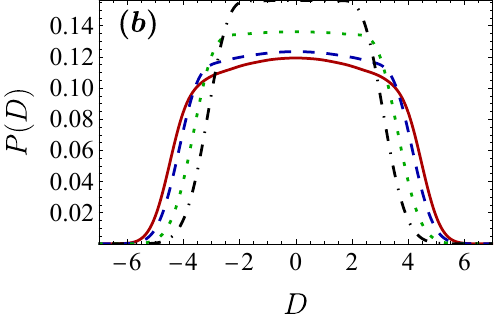}
\caption{Steady state detector probability $P(D)$ for (a) the Ising model for different system sizes $L$ with parameters $J=1$, $h=0.05$ and $\lambda=1$, $\gamma=2$ and (b) the LMG model for different values of $h=\{0.1,1.1,2.1,3.1\}$ (red solid, blue dashed, green dotted and black dot dashed lines) with parameters $L=4$, $\lambda=1$ and $\gamma=3$.} 
\label{fig:big}
\end{figure}

We can see the steady state detector probability $P(D)$ for different system sizes $L$ in Fig. \ref{fig:big}(a). Note that the peaks correspond to the eigenvalues of the measured operator $A$, which explains the structure and broadening of the distribution with increasing system size.

The Hamiltonian for the LMG model can be written in terms of the collective spin operators $S_q=\sum_{i=1}^L \sigma_i^q$ as
\begin{eqnarray}
    H= - \frac{1}{L} S_x^2+ h S_z.
\end{eqnarray}
For this model we will consider observable $A=S_y$.

In Fig. \ref{fig:big}(b), we can an example of the steady state $P(D)$ for the LMG model for $L=4$. Note that the large measurement bandwidth relative to the measurement strength leads to a less peaked distribution.

%%%%%%%%%%%%%%%%%%%%%%%%%%

\subsection{Ground state feedback stabilization}
In this final example, we will demonstrate the numerical advantage of the perturbative approach in calculating the feedback dynamics. The setup consists of a single qubit weakly in contact with a thermal bath, described by 
\begin{eqnarray}
    \mathcal{L}_0 = \kappa n_B \mathcal{D}[\sigma_+]+\kappa (n_B+1)\mathcal{D}[\sigma_-],
\end{eqnarray}
where $\kappa$ is the strength of the coupling and $n_B$ is set by the temperature. The goal is to use feedback control to maintain the qubit in the ground state by correcting for the effects of the bath.
The strategy \cite{feed1,feed2} is to measure $A=\sigma_x$ and apply a feedback protocol described by $f_1(D)=D$ and $\mathcal{L}_1= -i[g \sigma_y,\cdot]$. The intuition is that any measured deviation from the ground state is counteracted by this rotation.

\begin{figure}[t]
\centering
\includegraphics[width=0.45\linewidth]{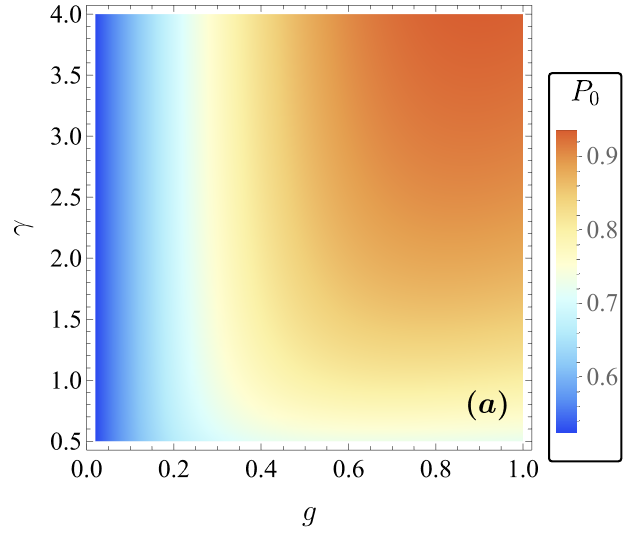}
\includegraphics[width=0.53\linewidth]{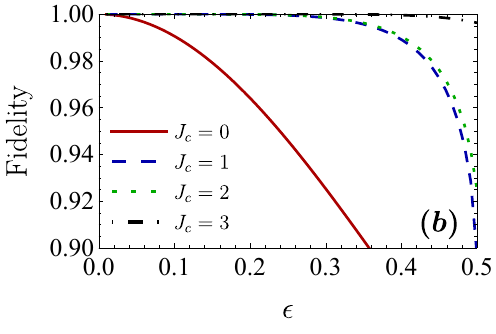}
\caption{Ground state feedback stabilization with measurement strength $\lambda=0.5$, coupling $\kappa=0.01$ and temperature $n_B=0.5$. (a) Steady state probability for the ground state against feedback strength $g$ and measurement bandwidth $\gamma$. (b)  Fidelity between unconditional steady states using exact solution and the perturbative solution with $J_c$ iterations against feedback strength $\epsilon$ and $\gamma=4$.} 
\label{fig:feedtest}
\end{figure}

In Fig. \ref{fig:feedtest}(a), we can see that the feedback control requires sufficient strength $g$ and a short response time $1/\gamma$ to successfully counteract thermalization due to the weakly coupled bath. This is characterized by the ground state probability $P_0=\bra{0}M_0 \ket{0}$ for the unconditional steady state. In the absence of feedback ($g=0$), the steady state (including the effects of measurement back action) has a ground state population $P_0\approx 0.505$, which agrees with the results shown.

% \begin{figure}[t]
% \centering
% \includegraphics[width=0.8\linewidth]{Feedbacktest.pdf}
% \caption{Steady state probability for the ground state against feedback strength $g$ and measurement bandwidth $\gamma$ with a measurement strength $\lambda=0.5$, coupling $\kappa=0.01$ and temperature $n_B=0.5$ } 
% \label{fig:feedtest}
% \end{figure}

We now show how our perturbative method can efficiently recreate these results, where $g$ plays the role of $\epsilon$. In Fig. \ref{fig:feedtest}(b) we compare the fidelity between the unconditional steady state obtained by exact diagonalization of the full matrix $\hat{Q}$ and using the perturbative method, which has a comparable computational speed up. Higher order corrections clearly provided increased fidelity with the exact steady state solution. For $\epsilon = 0.5$, the steady states with and without feedback already differ significantly, while the perturbative solution remains quite close to the exact result.

\section{Conclusion}

In this work, we studied feedback control in a continuously monitored quantum system subject to a low-pass filter, as recently derived in Ref.~\cite{annby-anderssonQuantumFokkerPlanckMaster2022}. We explore a solution method based on decomposing the signal-resolved density matrix in a series over Hermite polynomials, and showed how this can yield efficient access to various statistical properties of the signal, as well as measures of system-signal correlation. 
For the particular case of no feedback, we showed how the solution method simplifies dramatically, and can be implemented very efficiently. And we also illustrated how this could be used as the starting point for a perturbative solution involving weak feedback.

In the setting of open loop quantum control, there is a significant emphasis on both analytical techniques \cite{staReview} and numerically efficient methods to solve the dynamics which is used to systematically find optimal coherent control pulses \cite{Koch2016,Goerz2019}. Recent efforts in closed loop control have focused on machine learning to determine the optimal feedback control for a specific objective \cite{Borah2021,Porotti2022}. As experimental implementation of continuous weak measurements advances, our new approach will aid this with improved numerical efficiency and provide possible perturbative approaches to analytically determine the ideal feedback correction \cite{Whitty2020}.

% Going forward the block triangular structure of the dynamical matrix $\hat{Q}$ could also be used for fast computation of the dynamical evolution of the joint state.

% Going forward the block triangular structure of the dynamical matrix $\hat{Q}$ could also be used for fast computation of the dynamical evolution of the joint state i.e. by splitting the matrix into a block diagonal and nilpotent parts which commute. This could also be used for instance to compute two point correlation functions of the output current (see Appendix \ref{app3}) \anto{Does this make sense?! Maybe there are already better methods for this?}

\begin{acknowledgments}
\noindent{\textit{Acknowledgments.--}} AK acknowledges financial support of Taighde \'Eireann – Research Ireland under grant number 24/PATH-S/12701

\end{acknowledgments}

\begin{appendix}

\begin{widetext}

\section{Comparison with stochastic trajectories \label{app:comparison_stochastic_trajs}}

The standard way to describe a continuously monitored quantum system involves the use of stochastic master equations. For the setting described in the main text, this is the Belavkin equation,
\begin{eqnarray}\label{stochastic_eq_rho_c}
    d \rho_c = \mathcal{L}(D) \rho_c dt + \lambda \mathcal{D}[A] \rho_c dt + \sqrt{\lambda} \left\{ A-\avg{A}_c,\rho_c\right\} dW ,
\end{eqnarray}
where $\rho_c$ is the conditional density matrix, $dW$ is the Wiener increment and individual realizations of the measurement current obey the differential equation 
\begin{eqnarray}
    d D= \gamma \left(\avg{A}_c-D \right) dt + \frac{\gamma}{\sqrt{ 4\lambda}} dW,
\end{eqnarray}
with $\avg{A}_c=\Tr(A \rho_c)$. 
The stochastic unravelling provides a direct connection with experiments. And also allows for maximum flexibility, since one can implement any kind of feedback protocol desired: one simply modifies the Liouvillian $\mathcal{L}(D)$ to have any desired dependence on $D$; in a single stochastic trajectory, the obtained value of $D$ at each step is then simply input back into Eq.~\eqref{stochastic_eq_rho_c}. One can draw a direct analogy with the comparison between the Langevin equation and the Fokker-Planck equation in classical Brownian motion. 
Both equations describe the same physical phenomena, but from different angles. 
And therefore both have their uses. 
% It is challenging to give a fully general comparison between the numerical performance of the methods presented in the main text and that of stochastic trajectories (i.e. repeated simulations of the Belavkin equation). 

%%%%%%%%%%%%%%%%%%%%%%%%%%
\begin{figure}[h]
\centering
\includegraphics[width=\linewidth]{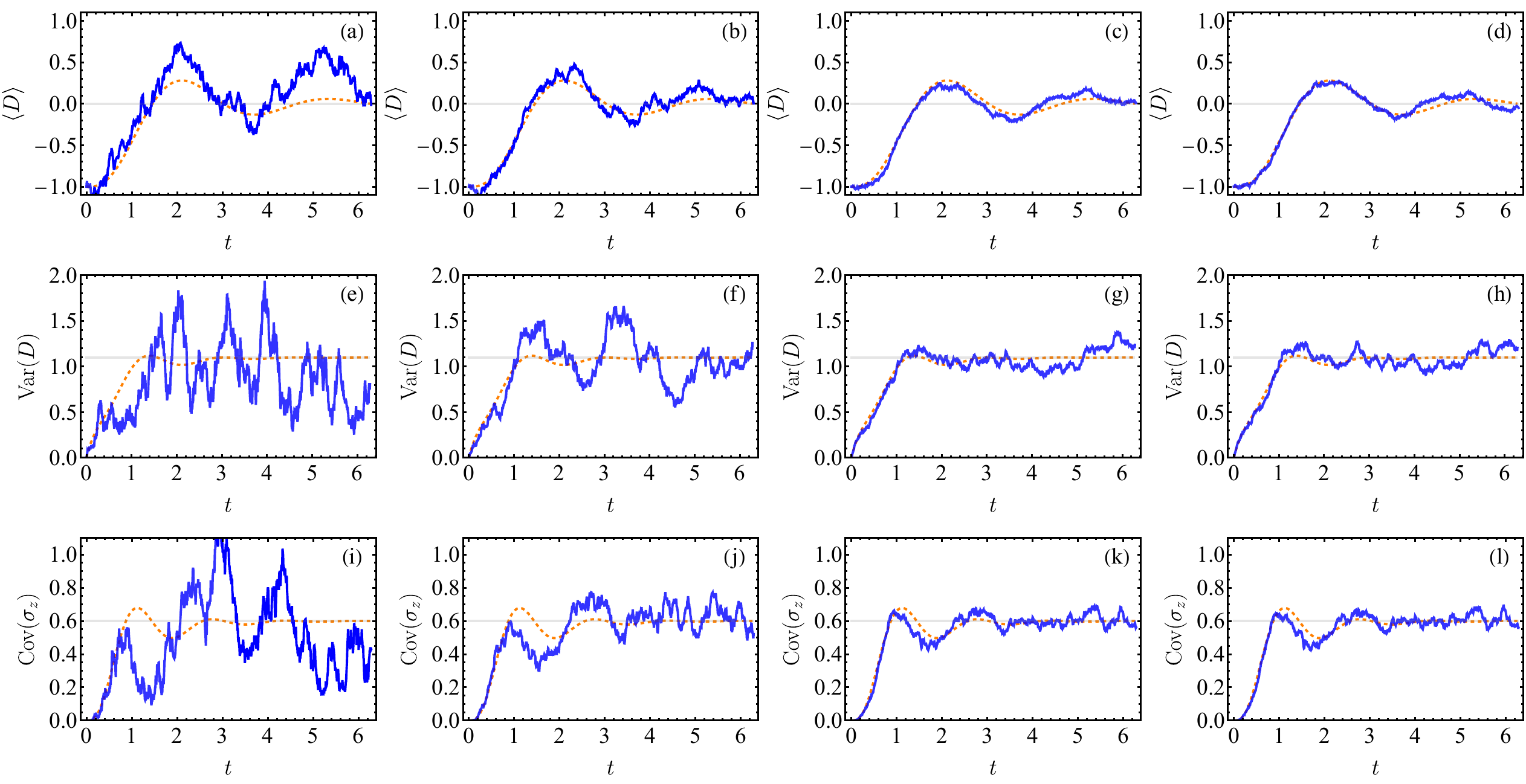} 
\caption{Average (top row) and variance (middle row) of filtered measurement current $D$, and covariance (bottom row) ${\rm Cov}(\sigma_z,D)$ using stochastic trajectories (blue solid line), dynamic Fokker-Planck solution (orange dashed line) and steady state Fokker-Planck solution (gray horizontal line). Number of stochastic trajectories are $10$ (a,e,i), $50$ (b,f,j),  $250$ (c,g,k),  $500$ (d,h,l). In all panels $N=5$ and other parameters are as in Fig. \ref{fig:compare}.} 
\label{fig:appendix}
\end{figure}
%%%%%%%%%%%%%%%%%%%%%%%%%%

We illustrate how stochastic trajectories compare with the deterministic equation~\eqref{FP_eq} through the lenses of a specific example. 
Namely, the same driven qubit example as in the main text (c.f. Fig. \ref{fig:compare}).
In Fig. \ref{fig:appendix} we show how the average, variance and covariance change in time for this process using both stochastic trajectories (blue) and Fokker-Planck equation solutions (orange). As noted in the main text, the steady state values for these parameters are exactly determined by our method to be $\avg{D}=0$, ${\rm Var}(D)=1.1$ and ${\rm Cov}(\sigma_z,D)=0.6$ (gray horizontal line). Note that while the first two quantities only depend on the statistics of the measurement current, the final quantity also depends on the conditional state.
The solution method proposed in the main text is particularly convenient for this problem, as it only required $N=5$ matrices in Eq.~\eqref{ansatz}. Moreover, our method goes directly to the steady-state; i.e., it does not require simulating transient effects, which is always necessary with stochastic trajectories. 
Fig. \ref{fig:appendix2} illustrate the steady-state mean and variance for different numbers of stochastic realizations. This provides an idea of how many samples are needed to reach convergence, although the precise value of course is problem-specific.

%%%%%%%%%%%%%%%%%%%%%%%%%%
\begin{figure}[h]
\centering
\includegraphics[width=0.65\linewidth]{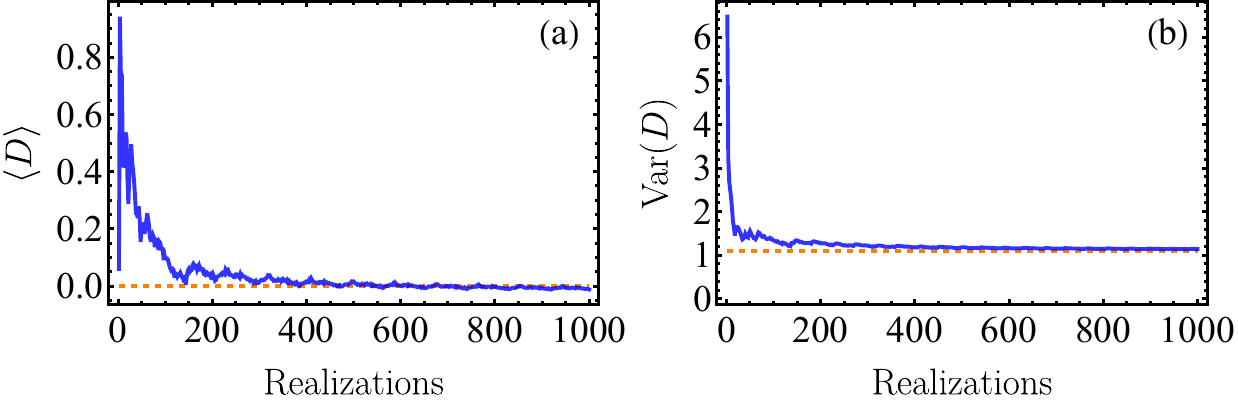} 
\caption{Steady state value of (a) the mean $\langle D \rangle$ and (b) the variance ${\rm Var}(D)$ against the number of stochastic realizations used (blue solid line). Exact steady state values are shown by orange dashed line and other parameters are as in Fig. \ref{fig:compare}.} 
\label{fig:appendix2}
\end{figure}
%%%%%%%%%%%%%%%%%%%%%%%%%%

\section{Informational quantities \label{app2}}

In the following, we outline some useful informational quantities that can be computed from the joint state.

\subsection{Mutual information between system and detector}
We consider the mutual information between the system and detector.  This can be expressed as
\begin{eqnarray}
\mathcal{I} = S \left[\int dD \rho_t(D) \right] + S \left[ \Tr\{\rho_t(D)\} \right]- S \left[ \rho_t(D) \right],
\end{eqnarray}
where $S \left[ \cdot \right]$ is either the von-Neumann or Shannon entropy depending on the input.  We simplify each term separately.  The first part is simply
\begin{eqnarray}
S \left[\int dD \rho_t(D) \right] &=& S[M_0] \\
&=& - \Tr M_0 \ln M_0.
\end{eqnarray}
The second term can be expanded as
\begin{eqnarray}
S \left[ \Tr\{\rho_t(D)\} \right] &=& S[P(D)] \\
&=& - \int dD P(D) \ln P(D).
\end{eqnarray}
The final term is then
\begin{eqnarray}
S \left[ \rho_t(D) \right] &=&- \int dD \, \Tr \left\{\rho_t(D) \ln \rho_t(D) \right\} .
\end{eqnarray}

By defining the density matrix given the outcome $D$ as $\tilde{\rho}_t(D)=\rho_t(D)/P(D)$ one can show that the mutual information is equivalent to
\begin{eqnarray}
    \mathcal{I}=-\int dD \, S\left[ \tilde{\rho}_t(D) \right] P(D)+S[M_0].
\end{eqnarray}
For a Hilbert space of dimension $R$, the maximum of this mutual information is then $\ln R$.

\subsection{Correlation}
If an observable has a spectral decomposition $B=\sum_n b_n \ketbra{b_n}{b_n}$, then there is a joint probability distribution $P(b_n,D)=\bra{b_n} \rho(D) \ket{b_n}$. One can then ask about the covariance of this distribution.

The expected measurement outcome is given by
\begin{eqnarray}
    \avg{D} &=& \int dD D P(D) \\
    &=&\sqrt{\sigma} c_1.
\end{eqnarray}
Similarly we have that for $B$, the unconditional expectation is
\begin{eqnarray}
    \avg{B} &=& \tr (M_0 B).
\end{eqnarray}
We define the covariance as
\begin{eqnarray}
    \rm{cov}(B,D) &=& \tr \int dD B D \rho(D) - \avg{D} \cdot \avg{B} \\
    &=& \sqrt{\sigma} \tr(M_1 B) - \sqrt{\sigma} c_1 \tr (M_0 B) \\
    &=& \sqrt{\sigma} \left[ \tr(M_1 B) -  \tr(M_1) \tr (M_0 B) \right].
\end{eqnarray}
This follows from the identity
\begin{eqnarray}
    \int dD D \frac{\mathcal{H}_n(D)}{\sqrt{\sigma^n n!}} \frac{e^{-D^2/2\sigma}}{\sqrt{2 \pi \sigma}} &=& J_n(1) \\
    &=& \sqrt{\sigma (n+1)} \delta_{n+1,0} + \sqrt{n \sigma} \delta_{n-1,0} \\
    &=&  \sqrt{n \sigma} \delta_{n-1,0},
\end{eqnarray}
% \begin{eqnarray}
%     \int dD D \frac{\mathcal{H}_n(D)}{\sqrt{\sigma^n n!}} \frac{e^{-D^2/2\sigma}}{\sqrt{2 \pi \sigma}} &=& \int dD  \frac{ \mathcal{H}_{n+1}(D)+n \sigma \mathcal{H}_{n-1}(D)}{\sqrt{\sigma^n n!}} \frac{e^{-D^2/2\sigma}}{\sqrt{2 \pi \sigma}} \\
%     &=& \sqrt{\sigma (n+1)} \delta_{n+1,0} + \sqrt{n \sigma} \delta_{n-1,0} \\
%     &=&  \sqrt{n \sigma} \delta_{n-1,0},
% \end{eqnarray}
since $n \geq 0$. Note that $\rm{Cov}(B)$ is identically zero if $M_1=0$, which is the case for a product state between system and detector.

\section{Two time correlation of filtered output current \label{app3}}
We want to compute $\avg{D(t)D(s)}$, which is explicitly given by
\begin{eqnarray}
    \avg{D(t)D(s)} = \int_{-\infty}^t \int_{-\infty}^s dt_2 dt_1 \gamma^2 e^{-\gamma (t-t_2+s-t_1)} \avg{z(t_1) z(t_2)}.
\end{eqnarray}
% We now calculate the term inside the integral using Eq. 72 of \cite{Landi2024}. We make the following correspondence $\nu=1/(2 \sqrt{\lambda})$, $\phi_k=0$, $L=\sqrt{\lambda}A$, $\mathcal{H}=1/2 \,\mathcal{C}_A$. From this we get that in the steady state,
The expression for the term inside the integral is given in steady state by \cite{Landi2024},
\begin{eqnarray}
    \avg{z(t)z(t+t')} &=&  \frac{1}{4 \lambda} \delta(t') + \frac{1}{4} \Tr \left[\mathcal{C}_A e^{\Lambda t'} \mathcal{C}_A M_0\right] \\
    &=&  \frac{1}{4 \lambda} \delta(t') + \Tr(A M_0)^2 + \frac{1}{2}\sum_{j \neq 0} e^{\eta_j t'} \Tr(A x_j) \Tr(y_j^\dagger \{A,M_0\}),
\end{eqnarray}
where we have used the spectral decomposition of $\Lambda$ and assumed no feedback. The end result is
% \begin{eqnarray}
%     \avg{D(t+\tau)D(t)} &=& \gamma^2 \int_{0}^{t+\tau} \int_{0}^t dt_2 dt_1 e^{-\gamma (2t+\tau-t_2-t_1)} \left\{ \frac{1}{4 \lambda} \delta(t_2-t_1) + \Tr(A M_0)^2 + \frac{1}{2}\sum_{j \neq 0} e^{\eta_j |t_2-t_1|} \Tr(A x_j) \Tr(y_j^\dagger \{A,M_0\}) \right\} \\
%     &=& \frac{\gamma  e^{-\gamma  (t+\tau )} \sinh (\gamma 
%    t)}{4 \lambda } + \left(e^{\gamma  t}-1\right) e^{-\gamma  (2 t+\tau )}
%    \left[e^{\gamma  (t+\tau )}-1\right] \Tr(A M_0)^2 \nonumber \\
%    &+& \sum_{j \neq 0}  \frac{\gamma  e^{-\gamma  (2 t+\tau )} \left(\eta _j
%    \left(e^{2 \gamma  t}-1\right)-\gamma 
%    \left(e^{(t+\tau ) \left(\gamma +\eta
%    _j\right)}-e^{\tau  \left(\gamma +\eta _j\right)+2
%    \gamma  t}+e^{t \left(\gamma +\eta
%    _j\right)}-1\right)\right)}{2\left(\gamma -\eta
%    _j\right) \left(\gamma +\eta _j\right)} \Tr(A x_j) \Tr(y_j^\dagger \{A,M_0\}).
% \end{eqnarray}
\begin{eqnarray}
    \avg{D(t+\tau)D(t)} &=& \gamma^2 \int_{-\infty}^{t+\tau} \int_{-\infty}^t dt_2 dt_1 e^{-\gamma (2t+\tau-t_2-t_1)} \left\{ \frac{1}{4 \lambda} \delta(t_2-t_1) + \Tr(A M_0)^2 + \frac{1}{2}\sum_{j \neq 0} e^{\eta_j |t_2-t_1|} \Tr(A x_j) \Tr(y_j^\dagger \{A,M_0\}) \right\} \\
   &=& \sigma e^{-\gamma \tau} + \Tr(A M_0)^2 + \frac{1}{2} \sum_{j \neq 0}  \frac{\gamma  \left(\gamma  e^{\tau  \eta
   _j}+e^{-\gamma  \tau } \eta _j\right)}{\gamma
   ^2-\eta _j^2} \Tr(A x_j) \Tr(y_j^\dagger \{A,M_0\}).
\end{eqnarray}
The $\tau=0$ case matches the result from the methods in the main text. Since the average filtered detector outcome in steady state is $\avg{D}=\Tr(A M_0)$, we can write our final result as
\begin{eqnarray}
C(\tau) &=& \avg{D(t+\tau)D(t)}-\avg{D(t)}^2\\
     &=&\sigma e^{-\gamma \tau} + \frac{1}{2} \sum_{j \neq 0}  \frac{\gamma  \left(\gamma  e^{\tau  \eta
   _j}+e^{-\gamma  \tau } \eta _j\right)}{\gamma
   ^2-\eta _j^2} \Tr(A x_j) \Tr(y_j^\dagger \{A,M_0\}).
\end{eqnarray}

\section{Fokker-Planck equation in Hermite polynomial basis \label{app}}

Since it will frequently arise,  we note the orthogonality relation
\begin{eqnarray}
\int_{-\infty}^\infty \frac{\mathcal{H}_m(x)}{\sqrt{\sigma^m m!}} \frac{\mathcal{H}_n(x)}{\sqrt{\sigma^n n!}} \frac{e^{-x^2/2\sigma}}{\sqrt{2 \pi \sigma}} dx
&=& \delta_{n,m}.
\end{eqnarray}
We start with the normalisation condition that $\int_{-\infty}^\infty dD \Tr\{\rho_t(D)\}=1$.  Using the orthogonality relation, we can write this as,
\begin{eqnarray}
\int_{-\infty}^\infty dD \, \Tr\{\rho_t(D)\} &=& \sum_{n=0}^{N-1} \Tr(M_n) \int_{-\infty}^\infty dD  \frac{\mathcal{H}_n(D)}{\sqrt{\sigma^n n!}} \frac{e^{-D^2/2\sigma}}{\sqrt{2 \pi \sigma}} \\
&=& \sum_{n=0}^{N-1} \Tr(M_n) \delta_{n,0} \\
&=& \Tr(M_0).
\end{eqnarray}
So the normalization condition simplifies to $\Tr(M_0)=1$, where $\int dD \rho_t(D) = \sum_n M_n \delta_{n,0}=M_0$ describes the state of the system when ignorant of the measurement outcome.

We now move on to the main equation.  Our strategy will be to insert our ansatz for the state, Eq. \eqref{ansatz} into the Eq. \eqref{FP_eq} and compute each term by applying $\int_{-\infty}^\infty dD \,\mathcal{H}_m(D)/\sqrt{\sigma^m m!}$ and utilizing the orthogonality relation (this is often indicated by an arrow $\rightarrow$).  

We start with the time derivative which only acts on the $M_n$ matrices leading to $\sum_{n=0}^{N-1} \dot{M}_n  \delta_{n,m}=\dot{M}_m$.  Next,  we will consider the first term on the right hand side of Eq.  \eqref{FP_eq}.  This is the outcome dependent Lindblad operator $\mathcal{L}(D)$.  This gives
\begin{eqnarray}
\sum_{n=0}^{N-1}   \int_{-\infty}^\infty dD \frac{\mathcal{H}_m(D)}{\sqrt{\sigma^m m!}}  \frac{\mathcal{H}_n(D)}{\sqrt{\sigma^n n!}} \frac{e^{-D^2/2\sigma}}{\sqrt{2 \pi \sigma}} \mathcal{L}(D)  M_n . \label{feed_term}
\end{eqnarray}
We write this in full generality as $\mathcal{L}(D)=\mathcal{L}_0+ \sum_p f_p(D) \mathcal{L}_p$. Altogether then Eq. \eqref{feed_term} simplifies to $\mathcal{L}_0 M_m + \sum_{n=0}^{N-1}  \sum_p \alpha_{n,m,p} \mathcal{L}_p M_n$ where
\begin{eqnarray}
\alpha_{n,m,p}(t)= \int_{-\infty}^\infty dD \frac{\mathcal{H}_m(D)}{\sqrt{\sigma^m m!}}  \frac{\mathcal{H}_n(D)}{\sqrt{\sigma^n n!}} \frac{e^{-D^2/2\sigma}}{\sqrt{2 \pi \sigma}}  f_p(D).
\end{eqnarray}

% For simplicity,  we will assume this only consists of Hamiltonian driving $\mathcal{L}(D)=-i [H_t(D),\cdot]$.  The Hamiltonian can be expanded as $H_t(D)=\sum_{p=0}^{L^2-1} f_{p,t}(D) B_p$ where $B_p$ are matrices which span the space of $L \times L$ Hermitian matrices and $f_{p,t}(D)$ specifies the choice of feedback driving. For notational convenience,  we will summarise this as
% \begin{eqnarray}
% \alpha_{n,m,p}(t)= \int_{-\infty}^\infty dD \frac{\mathcal{H}_m(D)}{\sqrt{\sigma^m m!}}  \frac{\mathcal{H}_n(D)}{\sqrt{\sigma^n n!}} \frac{e^{-D^2/2\sigma}}{\sqrt{2 \pi \sigma}}  f_{p,t}(D).
% \end{eqnarray}
% Altogether then Eq. \eqref{feed_term} simplifies to
% \begin{eqnarray}
% -i\sum_{n=0}^{N-1} \sum_{p=0}^{L^2-1}  [B_p,M_n] \alpha_{n,m,p}(t).
% \end{eqnarray}

% Other specific case is where some driving terms,  say $f_{1,t}(D)=g(t)$,  are not feedback dependant.  In this case the coefficient simplifies to $\alpha_{n,m,1}(t)= g(t) \delta_{n,m}$.  In general however,  the coefficients $\alpha_{n,m,p}(t)$ will depend on the choice of feedback driving $f_{p,t}(D)$. 

Moving on to the measurement back action term,  it simplifies as $-\frac{\lambda}{2}\sum_{n=0}^{N-1} \left[A,\left[A,M_n\right] \right] \delta_{n,m}=-\frac{\lambda}{2}\left[A,\left[A,M_m\right]\right]$ where  $[\cdot,\cdot]$ is the commutator. The second last term in Eq. \eqref{FP_eq} (describing the overall drift) is given by
\begin{eqnarray}
-\gamma \partial_D \mathcal{C}_{A-D} \rho_t(D)/2 &=& -\frac{\gamma}{2} \partial_D \{A-D,\rho_t(D)\} \\
&=& -\frac{\gamma}{2} \left[\{-1,\rho_t(D)\} + \{A-D,  \partial_D \rho_t(D)\} \right],
\end{eqnarray}
where we write the identity as $1$.  In order to evaluate $\partial_D \rho_t(D)$,  we note that
\begin{eqnarray}
\partial_D \left[ \mathcal{H}_n(D) e^{-D^2/2\sigma} \right]= -e^{-D^2/2\sigma} \sigma^{-1} \mathcal{H}_{n+1}(D).
\end{eqnarray}
Inserting the ansatz Eq.\eqref{ansatz} and applying $\int_{-\infty}^\infty dD \,\mathcal{H}_m(D)/\sqrt{\sigma^m m!}$ the drift term can be evaluated in separate parts as,
\begin{eqnarray}
\{-1,\rho_t(D)\}  &\rightarrow& \sum_n \{-1,M_n\} \delta_{n,m} \\
&=& \{-1,M_m\} \\
&=& - 2 M_m,
\end{eqnarray}
and
\begin{eqnarray}
\{A,  \partial_D \rho_t(D)\} &\rightarrow& -\sigma^{-1} \sum_n  \{A, M_n \} \delta_{n+1,m} \sqrt{(n+1)\sigma} \\
&=& -\sqrt{\frac{m}{\sigma}} \{A, M_{m-1} \}.
\end{eqnarray}
For the final part,  we will exploit the shift rule that $x \mathcal{H}_n(x)= \mathcal{H}_{n+1}(x)+ n \sigma \mathcal{H}_{n-1}(x)$. The last part is then
\begin{eqnarray}
\{-D,  \partial_D \rho_t(D)\} &=& - 2 D \, \partial_D \rho_t(D) \\
&=&  2 D \sigma^{-1} \sum_n M_n \frac{\mathcal{H}_{n+1}(D)}{\sqrt{\sigma^n n!}} \frac{e^{-D^2/2\sigma}}{\sqrt{2 \pi \sigma}} \\
&= &  2  \sigma^{-1} \sum_n M_n \frac{1}{\sqrt{\sigma^n n!}} \frac{e^{-D^2/2\sigma}}{\sqrt{2 \pi \sigma}} \left[\mathcal{H}_{n+2}(D)+ (n+1)\sigma \mathcal{H}_n(D) \right] \\
&\rightarrow & 2  \sigma^{-1} \sum_n M_n \left[ \delta_{n+2,m} \sqrt{(n+2)(n+1)} \sigma+(n+1)\sigma \delta_{n,m} \right] \\
&=& 2   \left[ \sqrt{m(m-1)} M_{m-2} + (m+1) M_m \right].
\end{eqnarray}
%\begin{eqnarray}
%\{-D,  \partial_D \rho_t(D)\} &\rightarrow& - \sum_n \{1, M_n\} \left( \sqrt{(n+1)\sigma} F_{m,n+1} + \frac{n \sigma}{\sqrt{n \sigma}} F_{m,n-1} \right) \\
%&=& - 2 \left[ \sqrt{m \sigma} M_{m-1} + \sqrt{(m+1) \sigma} M_{m+1} \right].
%\end{eqnarray}

Finally,  then we tackle the very last term in Eq. \eqref{FP_eq},  which can be thought of as a diffusion.  It can then be expressed as
\begin{eqnarray}
\frac{\gamma^2}{8 \lambda} \partial_D^2 \rho_t(D) &\rightarrow& \frac{\gamma^2}{8 \lambda} \frac{1}{\sigma^2} \sum_n M_n \delta_{n+2,m} \sqrt{(n+2)(n+1)} \sigma \\
&=& \frac{\gamma^2}{8 \lambda} \frac{\sqrt{m(m-1)}}{\sigma} M_{m-2} \\
&=& \gamma \sqrt{m(m-1)} M_{m-2}.
\end{eqnarray}

In complete summary,  the equation can then be written as the coupled differential equation
\begin{eqnarray}
\dot{M}_m = \mathcal{L}_0 M_m + \sum_{n=0}^{N-1}  \sum_p \alpha_{n,m,p} \mathcal{L}_p M_n -\frac{\lambda}{2}\left[A,\left[A,M_m\right]\right] -\frac{\gamma}{2} \left[ 2 m M_m-\sqrt{\frac{m}{\sigma}}   \{A,M_{m-1}\}  \right],
\end{eqnarray}
for each $m$ with the normalization condition $\Tr(M_0)=1$.  Note that $M_{-1}=0$.

\section{Matrix coefficients for polynomial feedback \label{app0}}
We wish to compute the coefficients $\alpha_{n,m,p}$ for a polynomial feedback function $f_p(D)=D^p$. This can be computed by noting that $\alpha_{n,m,0}=\delta_{n,m}$ and the recursive relation for this case
\begin{eqnarray}
    \alpha_{n,m,p} &=& \sqrt{\sigma (m+1)} \alpha_{n,m+1,p-1} + \sqrt{\sigma m} \alpha_{n,m-1,p-1}.
\end{eqnarray}
For example, the quadratic case is
\begin{eqnarray}
    \alpha_{n,m,2} &=& \sigma \sqrt{(m+1) (n+1)} \delta_{m+1,n+1} + \sigma \sqrt{n(m+1)} \delta_{m+1,n-1}+  \sigma \sqrt{m(n+1)} \delta_{n+1,m-1} + \sigma \sqrt{n m}  \delta_{n-1,m-1}.
\end{eqnarray}

\end{widetext}

\end{appendix}

\bibliography{ref}

\begin{thebibliography}{26}%
\makeatletter
\providecommand \@ifxundefined [1]{%
 \@ifx{#1\undefined}
}%
\providecommand \@ifnum [1]{%
 \ifnum #1\expandafter \@firstoftwo
 \else \expandafter \@secondoftwo
 \fi
}%
\providecommand \@ifx [1]{%
 \ifx #1\expandafter \@firstoftwo
 \else \expandafter \@secondoftwo
 \fi
}%
\providecommand \natexlab [1]{#1}%
\providecommand \enquote  [1]{``#1''}%
\providecommand \bibnamefont  [1]{#1}%
\providecommand \bibfnamefont [1]{#1}%
\providecommand \citenamefont [1]{#1}%
\providecommand \href@noop [0]{\@secondoftwo}%
\providecommand \href [0]{\begingroup \@sanitize@url \@href}%
\providecommand \@href[1]{\@@startlink{#1}\@@href}%
\providecommand \@@href[1]{\endgroup#1\@@endlink}%
\providecommand \@sanitize@url [0]{\catcode `\\12\catcode `\$12\catcode
  `\&12\catcode `\#12\catcode `\^12\catcode `\_12\catcode `\%12\relax}%
\providecommand \@@startlink[1]{}%
\providecommand \@@endlink[0]{}%
\providecommand \url  [0]{\begingroup\@sanitize@url \@url }%
\providecommand \@url [1]{\endgroup\@href {#1}{\urlprefix }}%
\providecommand \urlprefix  [0]{URL }%
\providecommand \Eprint [0]{\href }%
\providecommand \doibase [0]{http://dx.doi.org/}%
\providecommand \selectlanguage [0]{\@gobble}%
\providecommand \bibinfo  [0]{\@secondoftwo}%
\providecommand \bibfield  [0]{\@secondoftwo}%
\providecommand \translation [1]{[#1]}%
\providecommand \BibitemOpen [0]{}%
\providecommand \bibitemStop [0]{}%
\providecommand \bibitemNoStop [0]{.\EOS\space}%
\providecommand \EOS [0]{\spacefactor3000\relax}%
\providecommand \BibitemShut  [1]{\csname bibitem#1\endcsname}%
\let\auto@bib@innerbib\@empty
%</preamble>
\bibitem [{\citenamefont {Landi}\ \emph {et~al.}(2024)\citenamefont {Landi},
  \citenamefont {Kewming}, \citenamefont {Mitchison},\ and\ \citenamefont
  {Potts}}]{Landi2024}%
  \BibitemOpen
  \bibfield  {author} {\bibinfo {author} {\bibfnamefont {G.~T.}\ \bibnamefont
  {Landi}}, \bibinfo {author} {\bibfnamefont {M.~J.}\ \bibnamefont {Kewming}},
  \bibinfo {author} {\bibfnamefont {M.~T.}\ \bibnamefont {Mitchison}}, \ and\
  \bibinfo {author} {\bibfnamefont {P.~P.}\ \bibnamefont {Potts}},\ }\href
  {\doibase 10.1103/PRXQuantum.5.020201} {\bibfield  {journal} {\bibinfo
  {journal} {PRX Quantum}\ }\textbf {\bibinfo {volume} {5}},\ \bibinfo {pages}
  {020201} (\bibinfo {year} {2024})}\BibitemShut {NoStop}%
\bibitem [{\citenamefont {Gleyzes}\ \emph {et~al.}(2007)\citenamefont
  {Gleyzes}, \citenamefont {Kuhr}, \citenamefont {Guerlin}, \citenamefont
  {Bernu}, \citenamefont {Deleglise}, \citenamefont {Hoff}, \citenamefont
  {Brune}, \citenamefont {Raimond},\ and\ \citenamefont
  {Haroche}}]{gleyzes2007}%
  \BibitemOpen
  \bibfield  {author} {\bibinfo {author} {\bibfnamefont {S.}~\bibnamefont
  {Gleyzes}}, \bibinfo {author} {\bibfnamefont {S.}~\bibnamefont {Kuhr}},
  \bibinfo {author} {\bibfnamefont {C.}~\bibnamefont {Guerlin}}, \bibinfo
  {author} {\bibfnamefont {J.}~\bibnamefont {Bernu}}, \bibinfo {author}
  {\bibfnamefont {S.}~\bibnamefont {Deleglise}}, \bibinfo {author}
  {\bibfnamefont {U.~Busk}\ \bibnamefont {Hoff}}, \bibinfo {author}
  {\bibfnamefont {M.}~\bibnamefont {Brune}}, \bibinfo {author} {\bibfnamefont
  {J.M.}\ \bibnamefont {Raimond}}, \ and\ \bibinfo {author} {\bibfnamefont
  {S.}~\bibnamefont {Haroche}},\ }\href {https://doi.org/10.1038/nature05589}
  {\bibfield  {journal} {\bibinfo  {journal} {Nature}\ }\textbf {\bibinfo
  {volume} {446}},\ \bibinfo {pages} {297} (\bibinfo {year}
  {2007})}\BibitemShut {NoStop}%
\bibitem [{\citenamefont {Gustavsson}\ \emph {et~al.}(2006)\citenamefont
  {Gustavsson}, \citenamefont {Leturcq}, \citenamefont {Simovi{\v{c}}},
  \citenamefont {Schleser}, \citenamefont {Ihn}, \citenamefont {Studerus},
  \citenamefont {Ensslin}, \citenamefont {Driscoll},\ and\ \citenamefont
  {Gossard}}]{gustavsson2006}%
  \BibitemOpen
  \bibfield  {author} {\bibinfo {author} {\bibfnamefont {S.}~\bibnamefont
  {Gustavsson}}, \bibinfo {author} {\bibfnamefont {R.}~\bibnamefont {Leturcq}},
  \bibinfo {author} {\bibfnamefont {B.}~\bibnamefont {Simovi{\v{c}}}}, \bibinfo
  {author} {\bibfnamefont {R.}~\bibnamefont {Schleser}}, \bibinfo {author}
  {\bibfnamefont {T.}~\bibnamefont {Ihn}}, \bibinfo {author} {\bibfnamefont
  {P.}~\bibnamefont {Studerus}}, \bibinfo {author} {\bibfnamefont
  {K.}~\bibnamefont {Ensslin}}, \bibinfo {author} {\bibfnamefont {D.C.}\
  \bibnamefont {Driscoll}}, \ and\ \bibinfo {author} {\bibfnamefont {A.C.}\
  \bibnamefont {Gossard}},\ }\href
  {https://link.aps.org/doi/10.1103/PhysRevLett.96.076605} {\bibfield
  {journal} {\bibinfo  {journal} {Phys. Rev. Lett.}\ }\textbf {\bibinfo
  {volume} {96}},\ \bibinfo {pages} {076605} (\bibinfo {year}
  {2006})}\BibitemShut {NoStop}%
\bibitem [{\citenamefont {Bayer}\ \emph {et~al.}(2025)\citenamefont {Bayer},
  \citenamefont {Brange}, \citenamefont {Schmidt}, \citenamefont {Wagner},
  \citenamefont {Rugeramigabo}, \citenamefont {Flindt},\ and\ \citenamefont
  {Haug}}]{bayer2025}%
  \BibitemOpen
  \bibfield  {author} {\bibinfo {author} {\bibfnamefont {J.C.}\ \bibnamefont
  {Bayer}}, \bibinfo {author} {\bibfnamefont {F.}~\bibnamefont {Brange}},
  \bibinfo {author} {\bibfnamefont {A.}~\bibnamefont {Schmidt}}, \bibinfo
  {author} {\bibfnamefont {T.}~\bibnamefont {Wagner}}, \bibinfo {author}
  {\bibfnamefont {E.P.}\ \bibnamefont {Rugeramigabo}}, \bibinfo {author}
  {\bibfnamefont {C.}~\bibnamefont {Flindt}}, \ and\ \bibinfo {author}
  {\bibfnamefont {R.J.}\ \bibnamefont {Haug}},\ }\href
  {https://link.aps.org/doi/10.1103/PhysRevLett.134.046303} {\bibfield
  {journal} {\bibinfo  {journal} {Phys. Rev. Lett.}\ }\textbf {\bibinfo
  {volume} {134}},\ \bibinfo {pages} {046303} (\bibinfo {year}
  {2025})}\BibitemShut {NoStop}%
\bibitem [{\citenamefont {Vijay}\ \emph {et~al.}(2011)\citenamefont {Vijay},
  \citenamefont {Slichter},\ and\ \citenamefont {Siddiqi}}]{vijay2011}%
  \BibitemOpen
  \bibfield  {author} {\bibinfo {author} {\bibfnamefont {R.}~\bibnamefont
  {Vijay}}, \bibinfo {author} {\bibfnamefont {D.H.}\ \bibnamefont {Slichter}},
  \ and\ \bibinfo {author} {\bibfnamefont {I.}~\bibnamefont {Siddiqi}},\ }\href
  {https://link.aps.org/doi/10.1103/PhysRevLett.106.110502} {\bibfield
  {journal} {\bibinfo  {journal} {Phys. Rev. Lett.}\ }\textbf {\bibinfo
  {volume} {106}},\ \bibinfo {pages} {110502} (\bibinfo {year}
  {2011})}\BibitemShut {NoStop}%
\bibitem [{\citenamefont {Murch}\ \emph {et~al.}(2013)\citenamefont {Murch},
  \citenamefont {Weber}, \citenamefont {Macklin},\ and\ \citenamefont
  {Siddiqi}}]{murch2013}%
  \BibitemOpen
  \bibfield  {author} {\bibinfo {author} {\bibfnamefont {K.W.}\ \bibnamefont
  {Murch}}, \bibinfo {author} {\bibfnamefont {S.J.}\ \bibnamefont {Weber}},
  \bibinfo {author} {\bibfnamefont {C.}~\bibnamefont {Macklin}}, \ and\
  \bibinfo {author} {\bibfnamefont {I.}~\bibnamefont {Siddiqi}},\ }\href
  {https://doi.org/10.1038/nature12539} {\bibfield  {journal} {\bibinfo
  {journal} {Nature}\ }\textbf {\bibinfo {volume} {502}},\ \bibinfo {pages}
  {211} (\bibinfo {year} {2013})}\BibitemShut {NoStop}%
\bibitem [{\citenamefont {Minev}\ \emph {et~al.}(2019)\citenamefont {Minev},
  \citenamefont {Mundhada}, \citenamefont {Shankar}, \citenamefont {Reinhold},
  \citenamefont {Guti{\'e}rrez-J{\'a}uregui}, \citenamefont {Schoelkopf},
  \citenamefont {Mirrahimi}, \citenamefont {Carmichael},\ and\ \citenamefont
  {Devoret}}]{minev2019}%
  \BibitemOpen
  \bibfield  {author} {\bibinfo {author} {\bibfnamefont {Z.K.}\ \bibnamefont
  {Minev}}, \bibinfo {author} {\bibfnamefont {S.O.}\ \bibnamefont {Mundhada}},
  \bibinfo {author} {\bibfnamefont {S.}~\bibnamefont {Shankar}}, \bibinfo
  {author} {\bibfnamefont {P.}~\bibnamefont {Reinhold}}, \bibinfo {author}
  {\bibfnamefont {R.}~\bibnamefont {Guti{\'e}rrez-J{\'a}uregui}}, \bibinfo
  {author} {\bibfnamefont {R.J.}\ \bibnamefont {Schoelkopf}}, \bibinfo {author}
  {\bibfnamefont {M.}~\bibnamefont {Mirrahimi}}, \bibinfo {author}
  {\bibfnamefont {H.J.}\ \bibnamefont {Carmichael}}, \ and\ \bibinfo {author}
  {\bibfnamefont {M.H.}\ \bibnamefont {Devoret}},\ }\href
  {https://doi.org/10.1038/s41586-019-1287-z} {\bibfield  {journal} {\bibinfo
  {journal} {Nature}\ }\textbf {\bibinfo {volume} {570}},\ \bibinfo {pages}
  {200} (\bibinfo {year} {2019})}\BibitemShut {NoStop}%
\bibitem [{\citenamefont {Wiseman}\ and\ \citenamefont
  {Milburn}(2009)}]{wiseman2009}%
  \BibitemOpen
  \bibfield  {author} {\bibinfo {author} {\bibfnamefont {H.~M.}\ \bibnamefont
  {Wiseman}}\ and\ \bibinfo {author} {\bibfnamefont {G.~J.}\ \bibnamefont
  {Milburn}},\ }\href@noop {} {\emph {\bibinfo {title} {Quantum Measurement and
  Control}}}\ (\bibinfo  {publisher} {Cambridge University Press},\ \bibinfo
  {address} {New York},\ \bibinfo {year} {2009})\BibitemShut {NoStop}%
\bibitem [{\citenamefont {Kewming}\ \emph {et~al.}(2024)\citenamefont
  {Kewming}, \citenamefont {Kiely}, \citenamefont {Campbell},\ and\
  \citenamefont {Landi}}]{Kewming2024}%
  \BibitemOpen
  \bibfield  {author} {\bibinfo {author} {\bibfnamefont {M.~J.}\ \bibnamefont
  {Kewming}}, \bibinfo {author} {\bibfnamefont {A.}~\bibnamefont {Kiely}},
  \bibinfo {author} {\bibfnamefont {S.}~\bibnamefont {Campbell}}, \ and\
  \bibinfo {author} {\bibfnamefont {G.~T.}\ \bibnamefont {Landi}},\ }\href
  {\doibase 10.1103/PhysRevA.109.L050202} {\bibfield  {journal} {\bibinfo
  {journal} {Phys. Rev. A}\ }\textbf {\bibinfo {volume} {109}},\ \bibinfo
  {pages} {L050202} (\bibinfo {year} {2024})}\BibitemShut {NoStop}%
\bibitem [{\citenamefont {Annby-Andersson}\ \emph {et~al.}(2022)\citenamefont
  {Annby-Andersson}, \citenamefont {Bakhshinezhad}, \citenamefont
  {Bhattacharyya}, \citenamefont {De~Sousa}, \citenamefont {Jarzynski},
  \citenamefont {Samuelsson},\ and\ \citenamefont
  {Potts}}]{annby-anderssonQuantumFokkerPlanckMaster2022}%
  \BibitemOpen
  \bibfield  {author} {\bibinfo {author} {\bibfnamefont {B.}~\bibnamefont
  {Annby-Andersson}}, \bibinfo {author} {\bibfnamefont {F.}~\bibnamefont
  {Bakhshinezhad}}, \bibinfo {author} {\bibfnamefont {D.}~\bibnamefont
  {Bhattacharyya}}, \bibinfo {author} {\bibfnamefont {G.}~\bibnamefont
  {De~Sousa}}, \bibinfo {author} {\bibfnamefont {C.}~\bibnamefont {Jarzynski}},
  \bibinfo {author} {\bibfnamefont {P.}~\bibnamefont {Samuelsson}}, \ and\
  \bibinfo {author} {\bibfnamefont {P~P.}\ \bibnamefont {Potts}},\ }\href
  {\doibase 10.1103/PhysRevLett.129.050401} {\bibfield  {journal} {\bibinfo
  {journal} {Phys. Rev. Lett.}\ }\textbf {\bibinfo {volume} {129}},\ \bibinfo
  {pages} {050401} (\bibinfo {year} {2022})}\BibitemShut {NoStop}%
\bibitem [{\citenamefont {Diotallevi}\ \emph {et~al.}(2024)\citenamefont
  {Diotallevi}, \citenamefont {Annby-Andersson}, \citenamefont {Samuelsson},
  \citenamefont {Tavakoli},\ and\ \citenamefont
  {Bakhshinezhad}}]{Diotallevi_2024}%
  \BibitemOpen
  \bibfield  {author} {\bibinfo {author} {\bibfnamefont {G.~F.}\ \bibnamefont
  {Diotallevi}}, \bibinfo {author} {\bibfnamefont {B.}~\bibnamefont
  {Annby-Andersson}}, \bibinfo {author} {\bibfnamefont {P.}~\bibnamefont
  {Samuelsson}}, \bibinfo {author} {\bibfnamefont {A.}~\bibnamefont
  {Tavakoli}}, \ and\ \bibinfo {author} {\bibfnamefont {P.}~\bibnamefont
  {Bakhshinezhad}},\ }\href {\doibase 10.1088/1367-2630/ad3f3d} {\bibfield
  {journal} {\bibinfo  {journal} {New J. Phys.}\ }\textbf {\bibinfo {volume}
  {26}},\ \bibinfo {pages} {053005} (\bibinfo {year} {2024})}\BibitemShut
  {NoStop}%
\bibitem [{\citenamefont {Sousa}\ \emph {et~al.}(2025)\citenamefont {Sousa},
  \citenamefont {Bakhshinezhad}, \citenamefont {Annby-Andersson}, \citenamefont
  {Samuelsson}, \citenamefont {Potts},\ and\ \citenamefont
  {Jarzynski}}]{desousa2024}%
  \BibitemOpen
  \bibfield  {author} {\bibinfo {author} {\bibfnamefont {G.~De}\ \bibnamefont
  {Sousa}}, \bibinfo {author} {\bibfnamefont {P.}~\bibnamefont
  {Bakhshinezhad}}, \bibinfo {author} {\bibfnamefont {B.}~\bibnamefont
  {Annby-Andersson}}, \bibinfo {author} {\bibfnamefont {P.}~\bibnamefont
  {Samuelsson}}, \bibinfo {author} {\bibfnamefont {P.P.}\ \bibnamefont
  {Potts}}, \ and\ \bibinfo {author} {\bibfnamefont {C.}~\bibnamefont
  {Jarzynski}},\ }\href {\doibase 10.1103/PhysRevE.111.014152} {\bibfield
  {journal} {\bibinfo  {journal} {Phys. Rev. E}\ }\textbf {\bibinfo {volume}
  {111}},\ \bibinfo {pages} {014152} (\bibinfo {year} {2025})}\BibitemShut
  {NoStop}%
\bibitem [{\citenamefont {Annby-Andersson}\ \emph {et~al.}(2024)\citenamefont
  {Annby-Andersson}, \citenamefont {Bhattacharyya}, \citenamefont
  {Bakhshinezhad}, \citenamefont {Holst}, \citenamefont {Sousa}, \citenamefont
  {Jarzynski}, \citenamefont {Samuelsson},\ and\ \citenamefont
  {Potts}}]{annby2024maxwell}%
  \BibitemOpen
  \bibfield  {author} {\bibinfo {author} {\bibfnamefont {B.}~\bibnamefont
  {Annby-Andersson}}, \bibinfo {author} {\bibfnamefont {D.}~\bibnamefont
  {Bhattacharyya}}, \bibinfo {author} {\bibfnamefont {P.}~\bibnamefont
  {Bakhshinezhad}}, \bibinfo {author} {\bibfnamefont {D.}~\bibnamefont
  {Holst}}, \bibinfo {author} {\bibfnamefont {G.~De}\ \bibnamefont {Sousa}},
  \bibinfo {author} {\bibfnamefont {C.}~\bibnamefont {Jarzynski}}, \bibinfo
  {author} {\bibfnamefont {P.}~\bibnamefont {Samuelsson}}, \ and\ \bibinfo
  {author} {\bibfnamefont {P.P.}\ \bibnamefont {Potts}},\ }\href {\doibase
  10.1103/PhysRevResearch.6.043216} {\bibfield  {journal} {\bibinfo  {journal}
  {Phys. Rev. Res.}\ }\textbf {\bibinfo {volume} {6}},\ \bibinfo {pages}
  {043216} (\bibinfo {year} {2024})}\BibitemShut {NoStop}%
\bibitem [{\citenamefont {Wolf}\ \emph {et~al.}(2008)\citenamefont {Wolf},
  \citenamefont {Verstraete}, \citenamefont {Hastings},\ and\ \citenamefont
  {Cirac}}]{Wolf2008}%
  \BibitemOpen
  \bibfield  {author} {\bibinfo {author} {\bibfnamefont {M.~M.}\ \bibnamefont
  {Wolf}}, \bibinfo {author} {\bibfnamefont {F.}~\bibnamefont {Verstraete}},
  \bibinfo {author} {\bibfnamefont {M.~B.}\ \bibnamefont {Hastings}}, \ and\
  \bibinfo {author} {\bibfnamefont {J.~I.}\ \bibnamefont {Cirac}},\ }\href
  {\doibase 10.1103/PhysRevLett.100.070502} {\bibfield  {journal} {\bibinfo
  {journal} {Phys. Rev. Lett.}\ }\textbf {\bibinfo {volume} {100}},\ \bibinfo
  {pages} {070502} (\bibinfo {year} {2008})}\BibitemShut {NoStop}%
\bibitem [{\citenamefont {Cram{\'e}r}(1946)}]{Cramer1946}%
  \BibitemOpen
  \bibfield  {author} {\bibinfo {author} {\bibfnamefont {H.}~\bibnamefont
  {Cram{\'e}r}},\ }\href@noop {} {\emph {\bibinfo {title} {Mathematical Methods
  of Statistics}}}\ (\bibinfo  {publisher} {Princeton University Press},\
  \bibinfo {address} {Princeton, NJ},\ \bibinfo {year} {1946})\BibitemShut
  {NoStop}%
\bibitem [{\citenamefont {Rao}(1945)}]{Rao1945}%
  \BibitemOpen
  \bibfield  {author} {\bibinfo {author} {\bibfnamefont {C.~R.}\ \bibnamefont
  {Rao}},\ }\href@noop {} {\bibfield  {journal} {\bibinfo  {journal} {Bull.
  Calcutta Math. Soc.}\ }\textbf {\bibinfo {volume} {37}},\ \bibinfo {pages}
  {81--91} (\bibinfo {year} {1945})}\BibitemShut {NoStop}%
\bibitem [{\citenamefont {Smiga}\ \emph {et~al.}(2023)\citenamefont {Smiga},
  \citenamefont {Radaelli}, \citenamefont {Binder},\ and\ \citenamefont
  {Landi}}]{smiga2023}%
  \BibitemOpen
  \bibfield  {author} {\bibinfo {author} {\bibfnamefont {J.~A.}\ \bibnamefont
  {Smiga}}, \bibinfo {author} {\bibfnamefont {M.}~\bibnamefont {Radaelli}},
  \bibinfo {author} {\bibfnamefont {F.~C.}\ \bibnamefont {Binder}}, \ and\
  \bibinfo {author} {\bibfnamefont {G.~T.}\ \bibnamefont {Landi}},\ }\href
  {\doibase 10.1103/PhysRevResearch.5.033150} {\bibfield  {journal} {\bibinfo
  {journal} {Phys. Rev. Research}\ }\textbf {\bibinfo {volume} {5}},\ \bibinfo
  {pages} {033150} (\bibinfo {year} {2023})}\BibitemShut {NoStop}%
\bibitem [{\citenamefont {Facchi}\ and\ \citenamefont
  {Pascazio}(2008)}]{Facchi2008}%
  \BibitemOpen
  \bibfield  {author} {\bibinfo {author} {\bibfnamefont {P.}~\bibnamefont
  {Facchi}}\ and\ \bibinfo {author} {\bibfnamefont {S.}~\bibnamefont
  {Pascazio}},\ }\href {\doibase 10.1088/1751-8113/41/49/493001} {\bibfield
  {journal} {\bibinfo  {journal} {J. Phys. A: Math. Theor.}\ }\textbf {\bibinfo
  {volume} {41}},\ \bibinfo {pages} {493001} (\bibinfo {year}
  {2008})}\BibitemShut {NoStop}%
\bibitem [{\citenamefont {Prech}\ \emph {et~al.}(2025)\citenamefont {Prech},
  \citenamefont {Aschwanden},\ and\ \citenamefont {Potts}}]{feed1}%
  \BibitemOpen
  \bibfield  {author} {\bibinfo {author} {\bibfnamefont {K.}~\bibnamefont
  {Prech}}, \bibinfo {author} {\bibfnamefont {J.}~\bibnamefont {Aschwanden}}, \
  and\ \bibinfo {author} {\bibfnamefont {P.~P.}\ \bibnamefont {Potts}},\
  }\href@noop {} {\bibfield  {journal} {\bibinfo  {journal} {arXiv preprint
  arXiv:2505.16615}\ } (\bibinfo {year} {2025})}\BibitemShut {NoStop}%
\bibitem [{\citenamefont {Elouard}\ \emph {et~al.}(2017)\citenamefont
  {Elouard}, \citenamefont {Herrera-Mart\'{\i}}, \citenamefont {Huard},\ and\
  \citenamefont {Auff\`eves}}]{feed2}%
  \BibitemOpen
  \bibfield  {author} {\bibinfo {author} {\bibfnamefont {C.}~\bibnamefont
  {Elouard}}, \bibinfo {author} {\bibfnamefont {D.}~\bibnamefont
  {Herrera-Mart\'{\i}}}, \bibinfo {author} {\bibfnamefont {B.}~\bibnamefont
  {Huard}}, \ and\ \bibinfo {author} {\bibfnamefont {A.}~\bibnamefont
  {Auff\`eves}},\ }\href {\doibase 10.1103/PhysRevLett.118.260603} {\bibfield
  {journal} {\bibinfo  {journal} {Phys. Rev. Lett.}\ }\textbf {\bibinfo
  {volume} {118}},\ \bibinfo {pages} {260603} (\bibinfo {year}
  {2017})}\BibitemShut {NoStop}%
\bibitem [{\citenamefont {Gu\'ery-Odelin}\ \emph {et~al.}(2019)\citenamefont
  {Gu\'ery-Odelin}, \citenamefont {Ruschhaupt}, \citenamefont {Kiely},
  \citenamefont {Torrontegui}, \citenamefont {Mart\'{\i}nez-Garaot},\ and\
  \citenamefont {Muga}}]{staReview}%
  \BibitemOpen
  \bibfield  {author} {\bibinfo {author} {\bibfnamefont {D.}~\bibnamefont
  {Gu\'ery-Odelin}}, \bibinfo {author} {\bibfnamefont {A.}~\bibnamefont
  {Ruschhaupt}}, \bibinfo {author} {\bibfnamefont {A.}~\bibnamefont {Kiely}},
  \bibinfo {author} {\bibfnamefont {E.}~\bibnamefont {Torrontegui}}, \bibinfo
  {author} {\bibfnamefont {S.}~\bibnamefont {Mart\'{\i}nez-Garaot}}, \ and\
  \bibinfo {author} {\bibfnamefont {J.~G.}\ \bibnamefont {Muga}},\ }\href
  {\doibase 10.1103/RevModPhys.91.045001} {\bibfield  {journal} {\bibinfo
  {journal} {Rev. Mod. Phys.}\ }\textbf {\bibinfo {volume} {91}},\ \bibinfo
  {pages} {045001} (\bibinfo {year} {2019})}\BibitemShut {NoStop}%
\bibitem [{\citenamefont {Koch}(2016)}]{Koch2016}%
  \BibitemOpen
  \bibfield  {author} {\bibinfo {author} {\bibfnamefont {C.~P.}\ \bibnamefont
  {Koch}},\ }\href {\doibase 10.1088/0953-8984/28/21/213001} {\bibfield
  {journal} {\bibinfo  {journal} {J. Phys.: Condens. Matter}\ }\textbf
  {\bibinfo {volume} {28}},\ \bibinfo {pages} {213001} (\bibinfo {year}
  {2016})}\BibitemShut {NoStop}%
\bibitem [{\citenamefont {Goerz}\ \emph {et~al.}(2019)\citenamefont {Goerz},
  \citenamefont {Basilewitsch}, \citenamefont {Gago-Encinas}, \citenamefont
  {Krauss}, \citenamefont {Horn}, \citenamefont {Reich},\ and\ \citenamefont
  {Koch}}]{Goerz2019}%
  \BibitemOpen
  \bibfield  {author} {\bibinfo {author} {\bibfnamefont {M.~H.}\ \bibnamefont
  {Goerz}}, \bibinfo {author} {\bibfnamefont {D.}~\bibnamefont {Basilewitsch}},
  \bibinfo {author} {\bibfnamefont {F.}~\bibnamefont {Gago-Encinas}}, \bibinfo
  {author} {\bibfnamefont {M.~G.}\ \bibnamefont {Krauss}}, \bibinfo {author}
  {\bibfnamefont {K.~P.}\ \bibnamefont {Horn}}, \bibinfo {author}
  {\bibfnamefont {D.~M.}\ \bibnamefont {Reich}}, \ and\ \bibinfo {author}
  {\bibfnamefont {C.~P.}\ \bibnamefont {Koch}},\ }\href {\doibase
  10.21468/SciPostPhys.7.6.080} {\bibfield  {journal} {\bibinfo  {journal}
  {SciPost Phys.}\ }\textbf {\bibinfo {volume} {7}},\ \bibinfo {pages} {080}
  (\bibinfo {year} {2019})}\BibitemShut {NoStop}%
\bibitem [{\citenamefont {Borah}\ \emph {et~al.}(2021)\citenamefont {Borah},
  \citenamefont {Sarma}, \citenamefont {Kewming}, \citenamefont {Milburn},\
  and\ \citenamefont {Twamley}}]{Borah2021}%
  \BibitemOpen
  \bibfield  {author} {\bibinfo {author} {\bibfnamefont {S.}~\bibnamefont
  {Borah}}, \bibinfo {author} {\bibfnamefont {B.}~\bibnamefont {Sarma}},
  \bibinfo {author} {\bibfnamefont {M.}~\bibnamefont {Kewming}}, \bibinfo
  {author} {\bibfnamefont {G.~J.}\ \bibnamefont {Milburn}}, \ and\ \bibinfo
  {author} {\bibfnamefont {J.}~\bibnamefont {Twamley}},\ }\href {\doibase
  10.1103/PhysRevLett.127.190403} {\bibfield  {journal} {\bibinfo  {journal}
  {Phys. Rev. Lett.}\ }\textbf {\bibinfo {volume} {127}},\ \bibinfo {pages}
  {190403} (\bibinfo {year} {2021})}\BibitemShut {NoStop}%
\bibitem [{\citenamefont {Porotti}\ \emph {et~al.}(2022)\citenamefont
  {Porotti}, \citenamefont {Essig}, \citenamefont {Huard},\ and\ \citenamefont
  {Marquardt}}]{Porotti2022}%
  \BibitemOpen
  \bibfield  {author} {\bibinfo {author} {\bibfnamefont {R.}~\bibnamefont
  {Porotti}}, \bibinfo {author} {\bibfnamefont {A.}~\bibnamefont {Essig}},
  \bibinfo {author} {\bibfnamefont {B.}~\bibnamefont {Huard}}, \ and\ \bibinfo
  {author} {\bibfnamefont {F.}~\bibnamefont {Marquardt}},\ }\href {\doibase
  10.22331/q-2022-06-28-747} {\bibfield  {journal} {\bibinfo  {journal}
  {{Quantum}}\ }\textbf {\bibinfo {volume} {6}},\ \bibinfo {pages} {747}
  (\bibinfo {year} {2022})}\BibitemShut {NoStop}%
\bibitem [{\citenamefont {Whitty}\ \emph {et~al.}(2020)\citenamefont {Whitty},
  \citenamefont {Kiely},\ and\ \citenamefont {Ruschhaupt}}]{Whitty2020}%
  \BibitemOpen
  \bibfield  {author} {\bibinfo {author} {\bibfnamefont {C.}~\bibnamefont
  {Whitty}}, \bibinfo {author} {\bibfnamefont {A.}~\bibnamefont {Kiely}}, \
  and\ \bibinfo {author} {\bibfnamefont {A.}~\bibnamefont {Ruschhaupt}},\
  }\href {\doibase 10.1103/PhysRevResearch.2.023360} {\bibfield  {journal}
  {\bibinfo  {journal} {Phys. Rev. Res.}\ }\textbf {\bibinfo {volume} {2}},\
  \bibinfo {pages} {023360} (\bibinfo {year} {2020})}\BibitemShut {NoStop}%
\end{thebibliography}%
\end{document}